\documentclass[10pt, a4paper ]{article}

\usepackage[a4paper, total={6.7in, 9in}]{geometry}
\usepackage{setspace}
\usepackage{authblk}
\usepackage[numbers,sort&compress, super]{natbib}
\usepackage{mathtools}
\usepackage[justification=justified]{caption}

\usepackage{amsmath}
\usepackage{amsthm, mathtools}
\usepackage{amsfonts}
\usepackage{multicol, blindtext}
\usepackage{amssymb}
\usepackage[shortlabels]{enumitem}
\usepackage{qcircuit}
\usepackage{array}
\usepackage[english]{babel}
\usepackage{algorithm}
\usepackage{dsfont}
\usepackage{color} 
\usepackage[framemethod=TikZ]{mdframed}
\usepackage{algorithm}
\usepackage{algpseudocode}
\usepackage{braket}
\usepackage{etoolbox}
\apptocmd{\sloppy}{\hbadness 10000\relax}{}{}
\usepackage[labeled]{multibib}
\newcites{S}{Supplementary Material References}
\usepackage{url}

\usepackage{tikz}
\usepackage{graphicx}
\usetikzlibrary{positioning}
\usepackage[breaklinks=true,colorlinks=true,linkcolor=cyan,urlcolor=cyan,citecolor=blue]{hyperref}
\usepackage{wrapfig}
\usepackage{lineno} 
\usepackage{lipsum}
\usepackage{titlesec}

\usepackage[dvipsnames]{xcolor}
\usepackage{fontawesome5}

\usetikzlibrary{decorations.markings}
\tikzset{crossed/.style={
  decoration={markings, mark=at position 0.5 with {\draw[-] (-4pt,-4pt) -- (4pt,4pt);}},
  postaction={decorate}
}}

\usepackage{tcolorbox}
\newcommand{\mc}[1]{\mathcal{#1}}

\newcommand{\mf}[1]{\mathsf{#1}}

\newcommand{\kg}{\mathsf{KeyGen}}

\newcommand{\mac}{\mathsf{MAC}}

\newcommand{\macsign}{\mathsf{Mac}}
\newcommand{\macvfy}{\mathsf{Vfy}}

\newcommand{\kemenc}{\mathsf{Enc}}
\newcommand{\kemdec}{\mathsf{Dec}}

\newcommand{\ZZ}{\mathds{Z}}

\theoremstyle{definition}

\theoremstyle{plain}

\theoremstyle{remark}%

\newtheorem*{thm*}{Theorem} 

\newcommand{\getsr}{\stackrel{\$}{\gets}}

\usepackage{tikz}

\titleformat{\section}
  {\normalfont\large\bfseries}
  {\thesection}
  {0.7em}
  {}

\titleformat{\subsection}
  {\normalfont\normalsize\bfseries}
  {\thesubsection}
  {0.7em}
  {}

\newtheorem*{theorem*}{Theorem}

\begin{document}

\title{Authentication in Quantum Networks}
 

\author[1]{Christopher Battarbee}
\author[1]{Suchetana Goswami}
\author[1,2]{Elham Kashefi}
\author[1]{Mina Doosti}

\affil[1]{Quantum Software Lab, School of Informatics, University of Edinburgh, United Kingdom.}
\affil[2]{LIP6, CNRS, Sorbonne Université, Paris, France}


\date{}

\maketitle

\newcommand{\mina}[1]{\textcolor{blue}{[MD: #1]}}
\newcommand{\elham}[1]{\textcolor{orange}{[EK: #1]}}
\newcommand{\kit}[1]{\textcolor{teal}{[KB: #1]}}
\definecolor{chamoisee}{rgb}{0.6, 0.3, 0.3}
\newcommand{\suchetana}[1]{{\leavevmode\color{chamoisee}#1}}

\begin{abstract}
In this review, we survey the cryptographic task of authentication from the perspective of quantum communication. We review three main flavours of authentication that are often conflated in the literature: authentication of classical messages, authentication of quantum messages, and entity authentication, also covering recent hardware-assisted approaches. We compare representative protocols for each functionality in terms of their security assumptions, set-up requirements, composability, and scalability in large or dynamic networks, and use these criteria to identify and recommend suitable candidates. Finally, applications are surveyed: we provide a detailed case study of authentication and quantum key distribution (QKD), then extend the discussion to protocols beyond QKD, where the role of authentication is more complex. Our take-home message is that an authentication requirement is not an intrinsic limitation of quantum networks: as with all secure communication, each protocol relies on a particular authentication resource, and the security claim of that protocol is meaningful only once the authentication resource and its deployment assumptions are made explicit. At the same time, the existing classical and quantum literature already offers a range of quantum-secure authentication schemes, which can support different applications when carefully matched to the required functionality, assumptions, and security guarantees.
\end{abstract}

\begin{multicols}{2}

\section{Introduction}

Authentication is the dual problem of binding data to an identity, and ensuring that this data was not tampered with in transit. As a general rule, communication in the presence of an active adversary (or `man-in-the-middle') cannot hope to achieve any security, unless there is some means of providing authenticity. Quantum communication, meanwhile, can be regarded as the process of transmitting quantum data, and exploiting quantum mechanical properties to achieve security properties that are not possible with only classical information. 

Quantum information is difficult to deterministically control -- two measurements of the same quantum state can yield different results. Quantum communication therefore typically requires some classical discussion of the measurement process to accomplish its security goals, thereby posing a problem: how can we ensure the authenticity of the communication needed to establish security via quantum information?

A pertinent case study of this tension can be found in the debate surrounding QKD, a famous key-establishment protocol first proposed by Bennett and Brassard \cite{bb84_original}. In BB84, Alice prepares one of four polarisation states drawn from two conjugate bases. Bob independently chooses one of the two bases for each measurement. After transmission, Alice and Bob disclose their basis choices over an authenticated classical channel and retain the outcomes obtained in matching bases. They use a sample of these outcomes to estimate disturbance and, if the test succeeds, apply error correction and privacy amplification to distil a secret key. Security is void if an adversary can compromise the authenticity of the classical communication.

Various national security agencies consider this to be a fatal flaw.
Per the NSA's whitepaper\footnote{The British, French, German and Swedish national security agencies, among others, have produced similar documents that concur with this position.} on the subject \cite{nsa_position}: 
\begin{quote}
\textit{`QKD does not provide a means to authenticate the QKD transmission source. Therefore, source authentication requires the use of asymmetric cryptography or preplaced keys to provide that authentication. Moreover, the confidentiality services QKD offers can be provided by quantum-resistant cryptography, which is typically less expensive with a better understood risk profile.'}
\end{quote}


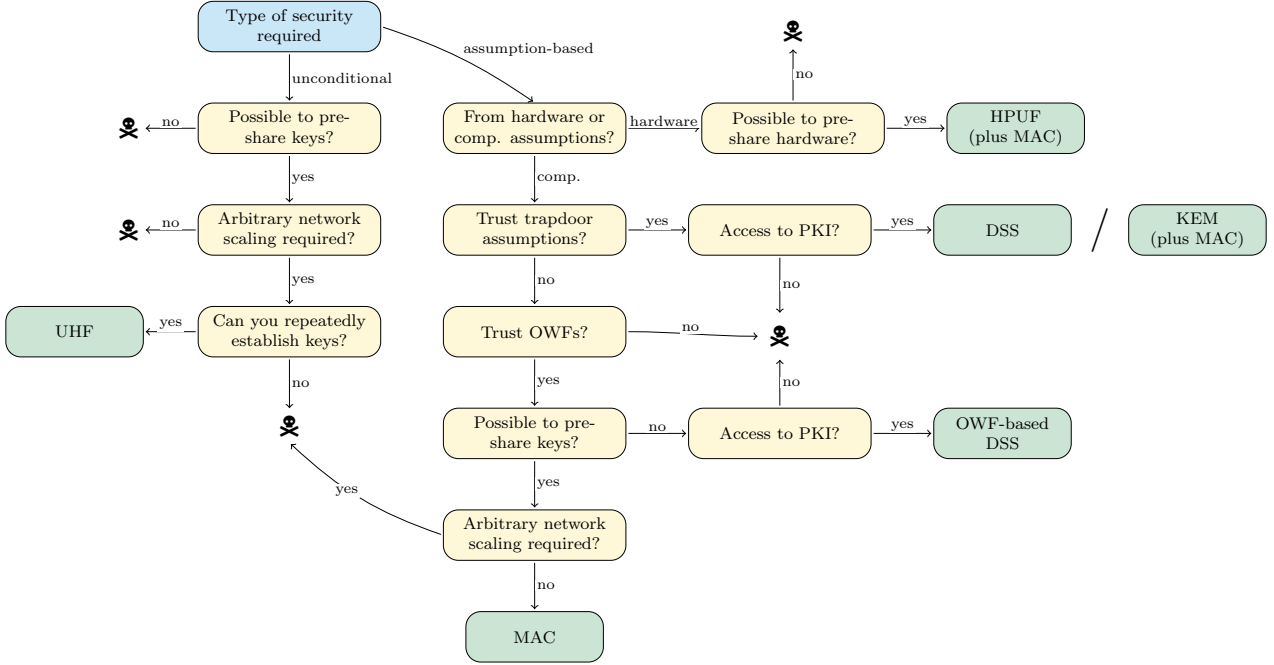
\begin{figure*}[!t]
    \centering
    \setlength{\abovecaptionskip}{4pt}%
    \setlength{\belowcaptionskip}{0pt}%
    \resizebox{0.98\textwidth}{!}{%
    \begin{tikzpicture}[
        node distance=1.32cm and 1.32cm,
        every node/.style={font=\footnotesize},
        start-box/.style={
            draw,
            rectangle,
            rounded corners=2.5mm,
            fill=Cerulean!20,
            minimum height=8.4mm,
            text width=2.85cm,
            align=center,
            inner sep=2.7pt
        },
        decision-box/.style={
            draw,
            rectangle,
            rounded corners=2.5mm,
            fill=Goldenrod!20,
            minimum height=8.4mm,
            text width=2.85cm,
            align=center,
            inner sep=2.7pt
        },
        end-box/.style={
            draw,
            rectangle,
            rounded corners=2.5mm,
            fill=ForestGreen!20,
            minimum height=8.4mm,
            text width=2.10cm,
            align=center,
            inner sep=2.7pt
        },
        no-go/.style={font=\normalsize, inner sep=2.5pt},
        arrow-label/.style={font=\scriptsize, fill=white, inner sep=1pt},
        flow-arrow/.style={->, shorten >=1pt, shorten <=1pt}
    ]

        \node[start-box] (start) {Type of security\\required};

        \node[decision-box, below=0.84cm of start] (d1) {Possible to pre-\\share keys?};
        \node[no-go, left=0.90cm of d1] (no-go-1) {\faSkullCrossbones};

        \node[decision-box, below=0.84cm of d1] (d2) {Arbitrary network\\scaling required?};
        \node[no-go, left=0.90cm of d2] (no-go-2) {\faSkullCrossbones};

        \node[decision-box, below=0.84cm of d2] (d3) {Can you repeatedly\\establish keys?};
        \node[end-box, left=0.90cm of d3] (uhf) {UHF};

        \node[no-go, below=0.88cm of d3] (no-go-3) {\faSkullCrossbones};

        \node[decision-box, right=1.03cm of d1] (d4) {From hardware or\\ comp. assumptions?};

        \node[decision-box, below=0.84cm of d4] (d5) {Trust trapdoor\\assumptions?};
        \node[decision-box, below=0.84cm of d5] (d6) {Trust OWFs?};
        \node[decision-box, below=0.84cm of d6] (d7) {Possible to pre-\\share keys?};
        \node[decision-box, below=0.84cm of d7] (d8) {Arbitrary network\\scaling required?};
        \node[end-box, below=0.84cm of d8] (mac) {MAC};

        \node[decision-box, right=1.03cm of d7] (d9) {Access to PKI?};
        \node[end-box, right=1.03cm of d9] (owf-dss) {OWF-based\\DSS};

        \node[no-go, above=0.92cm of d9] (no-go-6) {\faSkullCrossbones};

        \node[decision-box, right=1.03cm of d5] (d10) {Access to PKI?};

        \node[decision-box, right=1.25cm of d4] (d11) {Possible to pre-\\share hardware?};
        \node[no-go, above=0.92cm of d11] (no-go-4) {\faSkullCrossbones};
        \node[end-box, right=1.03cm of d11] (hpufs) {HPUF\\(plus MAC)};

        \node[end-box, right=1.03cm of d10] (dss) {DSS};
        \node[right=0.18cm of dss] (slash) {\huge /};
        \node[end-box, right=0.18cm of slash] (kem) {KEM\\(plus MAC)};

        \draw[flow-arrow] (start) -- (d1) node[arrow-label, midway, right] {unconditional};
        \draw[flow-arrow] (start.east) to[bend left=8]
            node[arrow-label, midway, above right] {assumption-based} (d4.north);

        \draw[flow-arrow] (d1) -- (no-go-1) node[arrow-label, midway, above] {no};
        \draw[flow-arrow] (d1) -- (d2) node[arrow-label, midway, right] {yes};
        \draw[flow-arrow] (d2) -- (no-go-2) node[arrow-label, midway, above] {no};
        \draw[flow-arrow] (d2) -- (d3) node[arrow-label, midway, right] {yes};
        \draw[flow-arrow] (d3) -- (uhf) node[arrow-label, midway, above] {yes};
        \draw[flow-arrow] (d3) -- (no-go-3) node[arrow-label, midway, right] {no};

        \draw[flow-arrow] (d4) -- (d5) node[arrow-label, midway, right] {comp.};
        \draw[flow-arrow] (d4) -- (d11) node[arrow-label, midway, above] {hardware};
        \draw[flow-arrow] (d11) -- (hpufs) node[arrow-label, midway, above] {yes};
        \draw[flow-arrow] (d11) -- (no-go-4) node[arrow-label, midway, right] {no};

        \draw[flow-arrow] (d5) -- (d6) node[arrow-label, midway, right] {no};
        \draw[flow-arrow] (d6) -- (d7) node[arrow-label, midway, right] {yes};
        \draw[flow-arrow] (d7) -- (d8) node[arrow-label, midway, right] {yes};
        \draw[flow-arrow] (d8) -- (mac) node[arrow-label, midway, right] {no};

        \draw[flow-arrow] (d8.west) to[bend left=12]
            node[arrow-label, midway, above left] {yes} (no-go-3.south);

        \draw[flow-arrow] (d5) -- (d10) node[arrow-label, midway, above] {yes};
        \draw[flow-arrow] (d10) -- (dss) node[arrow-label, midway, above] {yes};
        \draw[flow-arrow] (d10.south) -- ([yshift=2.5pt]no-go-6.north)
            node[arrow-label, midway, right] {no};

        \draw[flow-arrow] (d6.east) to[out=0, in=180]
            node[arrow-label, pos=0.48, above] {no} ([xshift=-2.5pt]no-go-6.west);
        \draw[flow-arrow] (d7) -- (d9) node[arrow-label, midway, above] {no};
        \draw[flow-arrow] (d9.north) -- ([yshift=-2.5pt]no-go-6.south)
            node[arrow-label, midway, right] {no};
        \draw[flow-arrow] (d9) -- (owf-dss) node[arrow-label, midway, above] {yes};

    \end{tikzpicture}%
    }
    \caption{Deciding the appropriate mechanism for classical message authentication. Start at the top and proceed by answering the questions as they arise; encountering the \faSkullCrossbones\ symbol implies that no protocol can provide the selected properties.}
    \label{fig:auth-flow}
\end{figure*}

In this review, we pay particular attention to the phrase ``source authentication requires the use of asymmetric cryptography or preplaced keys to provide that authentication''. We argue that all authentication requires some degree of set-up, and examine the authentication requirements of various quantum cryptographic protocols by breaking the study of authentication down into three main functionalities: authentication of classical messages, authentication of quantum messages, and identity authentication. For mechanisms providing these functionalities, we examine the stringency of their set-up requirements, their suitability for deployment at scale, and their security profiles. Since our focus is on quantum networks, we are particularly interested in schemes that provide quantum security, in the sense of security against adversaries with quantum capabilities who may be present on, or even fully control, the communication channel. In this setting, the interplay between set-up assumptions and security guarantees becomes especially important as different authentication mechanisms rely on different initial resources, and these resources determine both the strength and the scope of the resulting security claim. Understanding the different forms of authentication, their underlying assumptions, and their relevant figures of merit, including security level, composability, and scalability, is therefore essential for selecting a suitable mechanism for a given quantum communication task.

One of the main goals of this review is to unpack this selection problem by comparing both established and more recent approaches to authentication in quantum networks. We conclude with our recommendation of the best-fit authentication mechanisms for a variety of use cases, with particular attention to authentication in QKD because of its central role in the quantum communication community. At the same time, our analysis goes beyond QKD and we argue that many quantum communication and cryptographic protocols impose their own authentication requirements, and that these requirements must be matched carefully to the functionality, assumptions, and security guarantees of the authentication mechanism used.

\subsection{Organisation of the Review}
In Section~\ref{sec:framework} we set out some common ideas in authentication, and introduce the figures of merit relative to which the mechanisms are to be qualitatively analysed. We then turn to the most important schemes from classical authentication in Section~\ref{sec:classical-message}, from quantum authentication in Section~\ref{sec:quantum-message}, and from identity authentication in Section~\ref{sec:identification}.  In Section~\ref{sec:apps-and-depls} we conduct a thorough review of the QKD literature to see what authentication has used in practice and then discuss how the same authentication taxonomy appears in advanced quantum-network functionalities. Our findings on the most appropriate means of providing authentication are summarised in Figure~\ref{fig:auth-flow}, which is discussed in more detail in Section~\ref{sec:outlook}, along with future directions of research.

\section{Framework}\label{sec:framework}
As a cryptographic functionality, authentication comes with several distinct properties~\footnote{Here we introduce these properties in the level of the functionality of authentication not protocols, in the spirit of the framework in~\cite{singh2023towards}, while noting that the protocols implementing this functionality can also inherit them.}. Some of these properties capture the core security guarantees that an authentication mechanism is expected to provide, while others are additional features that may be useful in particular settings but are not always required. In this section, we first review these properties informally and explain their operational meaning, especially from the perspective of quantum-network protocols. We then introduce the main figures of merit one should consider when choosing an authentication mechanism, such as the level of security, the required set-up assumptions, and the scalability of the scheme. This provides the framework for the later comparison of the different approaches to building authentication.

\subsection{Properties of Authentication}\label{sec:props-of-auth}
\paragraph{Unforgeability.} The authentication of data is primarily accomplished in the \textit{message-tag paradigm}: to a message $m$ is appended a tag $t$, and there exists a \textit{verification algorithm} with Boolean output on any message-tag pair. \textit{Unforgeability} is then the notion that it should be difficult to come up with a pair $(m',t')$ that was not honestly generated by an authorised party, but is accepted by the verification algorithm. Formal definitions allow for various adversarial powers, such as access to honestly generated tags on messages of choice, or queries to the verification algorithm.

\paragraph{Integrity.} Related to unforgeability, this is the notion that precisely the information that was sent arrives at a receiver, and any modification to this data can be detected. Unforgeability generally provides integrity: if a message-tag pair $(m,t)$ is honestly generated, modification of this data will cause it to no longer be accepted by the verification algorithm. 

\paragraph{Identity misbinding.} Implicit to the idea of unforgeability is some \textit{a priori} calibration of the verification algorithm to accept message-tag pairs generated by the authorised party. If this calibration is off-set and the verification algorithm accepts data generated by an unauthorised party, then authenticity guarantees are void. This is accounted for in the set-up of authentication: for example, participants in a symmetric-key cryptosystem must be assured that only they possess the symmetric key.

\paragraph{Public verifiability.} It may or may not be desirable for the verification algorithm to be available to anyone; if it is, the authentication mechanism is said to be \textit{publicly verifiable}. We may also have a scenario where the verification algorithm can be shared with a limited number of parties, in which case the mechanism is \textit{transferably verifiable.} This is of particular interest in an asymmetric-key context, since this property enables public-key infrastructure. A dual notion is \textit{non-repudiation}; the inability of a party to convincingly deny having produced a particular message-tag pair. In the classical world, public verifiability prohibits repudiation; the situation is similar in the quantum world, but more subtle.

\paragraph{Non-malleability and controlled computation.} A fundamental aspect of quantum communication is that authentication of quantum data implies encryption; that is, any mechanism that prevents undetected tampering with unknown quantum data must also hide that data~\cite{barnum2002authentication}. As a result, the literature\cite{ambainis2009nonmalleable,alagic2017quantum} is interested in the related notion of \textit{non-malleability}, which says that operations on an encrypted message should not `usefully' affect the underlying message. Untangling the relationships between various types of quantum non-malleable encryption and quantum authentication is a task embarked upon in this review.

In this sense, non-malleability constrains adversarial attempts to induce meaningful changes to the plaintext, but it does not rule out transformations that are explicitly built into the functionality of a protocol.
Some protocols deliberately provide an evaluation interface, where prescribed gates or measurements can be applied to authenticated data while the
secret keys, traps, or syndromes are updated or checked. Such controlled computation is part of the intended functionality. The security
requirement is that operations outside this prescribed interface either act trivially, are independent of the input, or are detected. This
distinction is important in delegated-computation protocols, where authentication codes can be used not only to protect data, but also to
verify computation.

\subsection{Figures of Merit}
Authentication mechanisms have various desirable properties, but one cannot hope to achieve all of them simultaneously -- for example, higher security guarantees tend to come at the cost of more stringent set-up requirements. Here we list the figures of merit by which we `grade' the performance of the candidate authentication mechanisms.

\subsubsection{Security}
The most important measure of the suitability of a cryptographic mechanism is how well it provides its security guarantee. These guarantees come in two broad flavours.
\paragraph{Unconditional security.} Also known as \textit{information-theoretic} or \textit{statistical} security, a scheme is said to be \textit{unconditionally secure} if an attacker's success probability is bounded independently of its computational resources. The classic example of an unconditionally secure cryptosystem is the one-time pad: if a message is XORed with a uniformly random bitstring, any attacker cannot distinguish the resulting ciphertext from a uniformly random bitstring with probability greater than $1/2$, even if they are given unlimited time to do so. Typically achieving such a security level comes at a steep cost: for example, in the unconditionally secure variant of authentication (which we will go on to discuss), keys can only be used once before a new, uniformly random secret key is required.

\paragraph{Conditional security.} By assuming an attacker has some limited capabilities, we can relax some of our set-up assumptions. Most commonly, we assume that some computational problem is intractable, and that an attacker is sufficiently limited by its computational power to make brute-forcing this problem infeasible -- security achieved under these assumptions is called \textit{computational security}. To see the difference between this and unconditional security, contrast RSA to the one-time pad: if decryption is possible by finding some integer factorisation, then an attacker can tell the difference between a ciphertext and a random string with probability $1$ if factoring is easy, or if they are given unlimited resources with which to compute this factorisation -- whereas this is not possible for the one-time pad, regardless of the adversary's computational power.

Computational security can be further divided into the case where we assume the existence of \textit{one-way} functions, that is, functions that are difficult to invert given an output; and the existence of \textit{trapdoor} functions, that is, one-way functions that are generally difficult to invert unless one holds a `trapdoor' value. It is not known if there is separation between these two objects, but it is generally supposed\cite{five-worlds} that the existence of trapdoor functions is a stronger assumption than the existence of one-way functions. The terms of art here are Impagliazzo's hypothetical ``worlds'' \textit{minicrypt} and \textit{cryptomania}: if we live in \textit{minicrypt}, one-way functions exist and symmetric-key cryptography is possible; if we live in \textit{cryptomania}, trapdoor functions exist, and public-key cryptography is possible (we will return to what we mean by these terms in the next sub-section).

We can also make assumptions about physical devices. These are called \textit{hardware assumptions}: for example, we might assume that it is difficult to replicate random manufacturing defects. In practice this assumption does not necessarily hold classically due to various machine learning attacks -- as we will see, it is possible to augment these constructions with quantum technology. 

\paragraph{Quantum safety.} We are interested in security against classical adversaries, and adversaries with quantum computation capabilities. It is now a famous result that quantum algorithms can offer quadratic speed-up\cite{grovers-algorithm} in the analysis of symmetric-key cryptography, and exponential speed-up\cite{shors-algo} in the analysis of (at least) the public-key cryptography systems based on integer factorising and discrete logarithms. The currently deployed parameters for practical cryptography would require a much larger machine than we currently have available; nevertheless, recent (and rather dramatic) algorithmic improvements\cite{gidney-million-qubits}, as well as developments in \href{https://blog.google/innovation-and-ai/technology/research/google-willow-quantum-chip/}{hardware}, put the impending successful quantum cryptanalysis of classically secure cryptography firmly within the realm of plausibility.

In response to this growing threat, NIST\footnote{An American government standards agency, whose purview has traditionally included (but not monopolised) the establishment of cryptographic standards.} announced a call for alternative forms of public-key cryptography\cite{nist-pqc-announcement} believed not to be vulnerable to quantum cryptanalysis -- so-called \textit{post-quantum} cryptography. This process bore its first fruit in 2024 with the announcement of the first post-quantum standards\cite{fips203,fips204,fips205} -- in general these are much worse than their `pre-quantum' forebears from an efficiency standpoint, and so research into mitigating bandwidth costs, as well as developing advanced functionalities, is ongoing.

For our purposes, the main point of interest in post-quantum cryptography is in its interaction with quantum information, which is inherently quantum-safe (though quantum cryptography should certainly not be understood only as a tool for quantum-safe communication). As it turns out this interaction is rather subtle, and has been discussed in several existing reviews and commentaries\cite{alleaume_et_al_cryptographic_purposes,paterson_skepticism,the_case_for_qkd}; we provide our own perspective in a discussion section. 

It should be stressed that post-quantum cryptography is a subset of public-key cryptography, and so assuming its existence assumes the stronger cryptomania world. Moreover, its purported resistance to quantum cryptanalysis is an assumption supported by evidence, rather than an established fact. On the other hand, it is widely believed that traditional symmetric key cryptography is quantum-safe, since the best-known general quadratic speed-up in analysis\cite{grovers-algorithm} is not considered catastrophic (one can easily double parameter sets). In summary, if we live in minicrypt, then we believe that everything is quantum-resistant; if we live in cryptomania, algorithms may or may not be quantum-resistant, and more care is required.

\paragraph{Composability} A strong security property of a cryptographic mechanism is its \textit{composable} security; that is, it should be not only secure in a vacuum, but secure in composition with other cryptographic mechanisms. This is of particular interest to us, since we are considering cryptographic mechanisms that satisfy the assumed authentication requirements of quantum cryptographic protocols -- if these authentication mechanisms are composably secure then we have `plug-and-play' security without having to re-prove the security of the composite system.

In practice, composable security is generally achieved as a consequence of demonstrating the simulation-based security of a cryptographic mechanism; this is generally under-studied compared to the more prevalent and weaker notion of game-based security. We report throughout on the status of composable results surrounding each of the mechanisms we discuss. In fact, there are two major schools of thought on how best to demonstrate composable security: the Universal Composability (UC) framework\cite{canetti-uc} of Canetti, which has engendered many related frameworks generally used as proof tools in specialist scenarios; or the Constructive Cryptography (CC) framework of Maurer \cite{maurer-cc} (originally billed as \textit{Abstract} Cryptography, in a paper also co-authored with Renner). Despite being much less prevalent in the literature, the latter is better suited for our purposes, since it is generally preferred by quantum information theorists. As such, it will be more convenient for proofs of the composable security of our authentication mechanisms to go through in the CC framework, because then we will have easier to access to the `plug-and-play' security we desire.

For the classical setting, composability in the CC framework has recently been given an extensive treatment by one of the authors\cite{composable-review}, in a systematisation currently being prepared with this report. We will make frequent reference to these results.

\subsubsection{Set-up Requirements}
Consider a communication channel from an honest party $A$ to an honest party $B$ that is insecure in the following sense: all data input to the communication channel can be read by an adversary $E$, and all data received by party $B$ is selected by adversary $E$. This is a maximally insecure channel, offering no privacy or authenticity guarantees. On the other hand, it is in some sense `free' -- such a channel can always be realised, say by sending a typewritten letter, or communicating at the network layer of the internet. As such we consider this \textit{insecure channel} the basic building block of our communication, and assume that it is always available.

Clearly, we cannot hope for any authenticity guarantees with access only to an insecure channel, regardless of the authentication mechanism we apply to the data we send down this channel. We detail the typical additional set-up assumptions one can make, in roughly descending order of stringency.

\paragraph{Pre-shared classical keys.} This is the assumption that completely private (and usually uniformly randomly sampled) data is assumed to have been distributed ahead of time to the honest parties. If these keys are used once we can hope for unconditional security; if they are to be re-used we get the one-way function variant of computational security. 

Methods of distributing these pre-shared keys include meeting in person, or delivery by a trusted courier. In a survey discussing similar topics to ours\cite{paterson_skepticism}, the authors note that a possible solution is the distribution of pre-shared keys installed on hardware, citing contemporaneous widespread adoption of this approach with mobile SIM cards.

\paragraph{Pre-shared quantum keys.} Quantum digital signature schemes (confusingly, a method of authenticating classical data) require their public keys, which are quantum states, to arrive at the verifier unaltered. As such, in this case we actually have to assume quantum authentication to provide classical authentication. As we will see\cite{barnum2002authentication}, exotic properties of quantum information actually mandate that any scheme guaranteeing the authenticity of quantum information must also encrypt this quantum information, in a precisely defined sense. The requirement for quantum authentication therefore reduces to the requirement of classical or quantum pre-shared keys. Powerful functionalities can also be achieved if two parties possess each half of an entangled pair; indeed, \textit{a priori} access to entanglement is a common resource in the literature.

\paragraph{Pre-shared hardware.} Instead of pre-distributing classical strings we can pre-distribute hardware tokens. These are semantically similar to pre-shared keys (in that they each pair of communication partners must establish a distinct hardware token, and maintain its security), but with an interesting additional robustness: if the hardware token is compromised (say, physically intercepted), its security survives up to the validity of hardware assumptions. Contrast this to the compromise of pre-shared keys, which immediately voids all security guarantees.

\paragraph{Classical authentication.} Despite their goal of providing classical authentication, some of the schemes require classical authentication as an assumption. This primarily relates to \textit{public-key cryptography}; rather than distributing private keys, one of the parties generates a pair of keys, called the public and private keys; the public key can be shared arbitrarily, but leaking the private key voids the security guarantees. 

The point here is that honest participants must receive the \textit{correct} public key; if the adversary can inject their own choice of public key, then they know the corresponding secret key and security guarantees are void. As such the minimal assumption is an authenticated channel to enable faithful distribution of public keys (though in contrast to pre-shared key distribution, these public keys need not remain private, as the name suggests). 

\subsubsection{Scalability} 
A key measurement of how `expensive' we should consider the set-up requirements of a scheme is how these requirements scale with the number of participants in a network. A useful metric here is the difficulty a new participant has in joining the network, which we will refer to as \textit{joining cost}. 

\paragraph{Pre-shared keys.} The scaling problem inherent to networks of pre-shared keys is well-known; the basic problem is that each pair of participants must share a distinct pre-shared key, or any authenticity guarantees are void, because if party $A$ shares a key with both $B$ and $C$, it has no means of distinguishing which of these parties it is talking to\footnote{Unless we assume the honest behaviour of all parties in possession of a single pre-shared key, but this is quite a restrictive model.}. This means that $~n^2$ distinct keys in total need to be generated and distributed among the network; moreover, the joining cost is a function of the size of the network, since a new participant has to establish as many secret keys as there are network participants. 

One method of circumventing this issue is to introduce a trusted facilitating party -- this approach is the idea of a key-distribution centre (KDC), used perhaps most famously in the Kerberos protocol\cite{kereberos/kdc}. At the cost of complete trust in a central authority, parties can have pre-shared keys distributed to them by this authority, provided both parties share some long-term symmetric key with the central authority. In effect, this means that, by establishing a pre-shared key once, we can negotiate secret keys over insecure channels with an unbounded number of parties, provided those parties have also established a pre-shared key with the trusted authority; in other words, the joining cost is independent of the size of the network. 

A major drawback of the KDC approach is that the whole system is vulnerable to passive corruption of the trusted party, since reading its memory allows one to read all the communication it has facilitated. Moreover, we require the trusted authority to be online to facilitate key distribution, which limits scalability.

\paragraph{Hardware tokens.} The scalability profile of distributing hardware tokens is similar to pre-shared keys, in the sense that each pair of participants has to have exchanged a distinct hardware token. We note that the efficiency profile is often very asymmetric in hardware-based schemes: one party only has to reply to challenges by interacting with their hardware and sending responses, which is very cheap; but the other has to maintain a (possibly very large) set of challenge-response queries. As such, these are particularly well suited to star-shaped networks. 

\paragraph{Public keys.} As we have seen, distributing classical public keys has to assume some initial classical authentication, or security guarantees are void. In a network of $n$ participants, we immediately improve on the total number of keys in the network, because each participant has to distribute $n-1$ copies of its public keys (rather than $n-1$ distinct keys). On the other hand, the joining cost is still high, because a new participant has to establish $n$ authenticated channels with each of the existing participants to distribute their public key. Moreover, whether or not establishing a pre-shared key is by itself more difficult than establishing one-time authentication is a topic for debate -- for example, the most obvious method of establishing an initial one-time authentication is via an in-person meeting, at which pre-shared key distribution could also be accomplished. 

In order to solve this problem, we can proceed similarly to the KDC method. The idea is that one establishes one-time authentication with a trusted authority, this time called a Certificate Authority (CA), both authentically transmitting one's own public key, and receiving an authentic copy of the CA's public key. Because of the public verifiability of \textit{digital signature schemes}, which we will go on to discuss, we can now outsource to the CA the authentic distribution of public keys to peers by distributing a signature received from the CA. We will see how this is done in Section~\ref{sec:public}.

Even if the one-time joining cost for a CA is not necessarily better than in the case of a KDC we make two improvements compared to a KDC: the first is that passive compromise of the CA is no longer sufficient, and even an active compromise does not affect prior communication. The second is that scalability improves: the CA no longer has to be online to facilitate the distribution of public keys, and the certificates it distributes can be thought of a token that suffice for establishing authentication indefinitely. Crucially, \textit{CAs can register each other}, so that end users need not necessarily have registered with the same CA to allow authentic distribution of their public keys. Such ``chains of trust'' between several CAs are referred to as a Public Key Infrastructure (PKI).

We remark that the NSA's position on QKD (which we saw in the introduction) notes that ``QKD does not provide a means of authenticated the QKD transmission source'', and offers quantum-resistant cryptography as an alternative -- it is in this sense that precisely the same thing is true of quantum-resistant cryptography, which is just a sub-category of public key cryptography. The same objection has already been raised in slightly less precise language\cite{renner_wolf_rebuttal}.

We are now ready to begin the analysis of our three modes of authentication.

\section{Authentication of classical messages} \label{sec:classical-message}
Classical data broadcast over an insecure channel is considered authenticated if the data received from that channel is the same as the data input to that channel. The cryptography, however, does not guarantee that the correct party input data into the channel in the first place. Indeed, as we have alluded to in Section~\ref{sec:props-of-auth}, authentication is implicitly calibrated with respect to a particular party (or parties). We will return to discuss this issue once we have reviewed the means of achieving authentication.

\subsection{Symmetric-key methods} \label{sec:symmetric}
Supposing that both honest parties share a pre-shared private key, we discuss $3$ means of authentication.

\paragraph{Universal hash functions.} We can think of universal hash functions\footnote{The term `hash function' is overloaded in the literature; these objects are not to be confused with the more familiar \textit{cryptographic} hash functions, which take inputs of arbitrary length, and have altogether distinct security profiles.} (UHFs) as the `one-time pad of authentication'. Consider a family of functions $\mc{H}$ from a space of fixed-length messages $\mc{M}$, to a space of fixed-length tags $\mc{T}$. We can index this family of functions with a set of keys $\mc{K}$, so that every $k\in\mc{K}$ specifies a unique function $h_k\in\mc{H}$. Given a key $k\in\mc{K}$, tags are generated by computing $t\gets h_k(m)$; the verification algorithm corresponding to a key $k$ takes a message-tag pair $(m,t)$, and checks whether $h_k(m)=t$; outputting $\top$ if and only if this holds.

If this family of functions has a property called $\epsilon$-almost strong universality, introduced by Wegman and Carter\cite{wegman_carter_universal_hash,wegman_carter_authentication}, it turns out that we can achieve unconditionally secure message authentication. This is the following property: for every $m_1,m_2\in\mc{M}$ and $t_1,t_2\in\mc{T}$, we should have 

$$|\{h\in\mc{H}:h(m_1)=t_1\wedge h(m_2)=t_2\}|\leq\frac{\epsilon|\mc{H}|}{|\mc{T}|}$$

It turns out that taking $\epsilon=1/|\mc{T}|$ is optimal, in which case this property is also known as $2$-universality. Stinson\cite{stinson_universal_auth} shows that this property implies another property put forward by Wegman and Carter, namely that for each $m\in\mc{M}$ and $t\in\mc{T}$, we have 
\[|\{h\in\mc{H}:h(m)=t\}|\leq\frac{|\mc{H}|}{|\mc{T}|}\]
Taken together, these properties imply that seeing a single message-tag pair on an insecure channel leaks no information about which element of $\mc{H}$ was used to produce that pair, because each is equally likely; and that there is a small number distinct pairs that are valid under the same element of $\mc{H}$. Much later, Portmann would use these properties to demonstrate that universal hash functions yield composable, unconditionally secure message authentication\cite{portmann_key_recycling}. In particular, for $\epsilon$-almost strongly universal hash functions, given a valid message-tag pair, an adversary's probability of computing a valid tag on a fresh message is bounded by $\epsilon$. Note that the secret hash function used to compute tags must be selected uniformly at random for the security guarantees to hold. Equivalently, the secret key must be sampled uniformly at random.

The strong $2$-universality of these functions does not imply strong $3$-universality, and so on, so we only get security guarantees if a function $h$ (equivalently, a pre-shared key that specifies this function) is used once. In fact, it is known\cite{composable-review} that no family of functions can promise composable, unconditional security with respect to arbitrarily many uses. On the other hand, it is also known\cite{composable-review} that for one-time, composable, unconditional security, the pre-shared keys need to be roughly logarithmic in the size of the message. Compare this state of affairs favourably to the case of the one-time pad, whereby the keys have to be at least as large as the message. It is this gap that, in part, yields the quantum advantage in quantum key expansion, an approach that allows us to derive an arbitrary amount of unconditionally secret material from an initial pre-shared key. Even with quantum technology, this would not be possible if the classical transmission in QKD needed to be private.

\begin{tcolorbox}[colback=blue!10,colframe=blue!50, title=Wegman-Carter UHFs]
    Suppose we wish to compute tags on messages from a space $\mc{M}$ to a tag space $\mc{T}$. Set $n=|\mc{T}|$ and pick a prime $p$ such that $p>|\mc{M}|$. Pick $a,b\in\ZZ_p$ uniformly at random, and compute tags for a message $m$ by computing
    \[t=((ax+b)\bmod p)\bmod n\]

    The space of keys is $\ZZ_p\times \ZZ_p$, and each secret key has length $2\log p$, and the construction achieves $\epsilon$-almost strong universality for a small constant $\epsilon$.
\end{tcolorbox}

A natural method of constructing these functions is with modular arithmetic, intuitively because this evenly distributes integers among a fixed number of classes. Indeed this underpins Wegman and Carter's original construction. Later, care was taken to construct examples of such functions that did not require modular arithmetic, which can be expensive. Examples include GHASH\cite{ghash1,ghash2,ghash3}, the construction underpinning UMAC\cite{umac} (which is not itself a universal hash function -- more on this in the next section), and Poly1305\cite{poly1305}.

\begin{tikzpicture}
  \node[
    draw,
    rounded corners=6pt,
    fill=yellow!20,
    inner sep=8pt,
    text width=0.75\columnwidth,
    align=left
  ] (box) {

    \vspace{6pt}

    \textbf{security} \dotfill\ unconditional\\
    \textbf{set-up} \dotfill\ one-use PSKs\\
    \textbf{composability} \dotfill\ known\\
    \textbf{public verifiability} \dotfill\ no\\
    \textbf{scalability} \dotfill\ very poor
  };

  \node[
    draw,
    rounded corners=3pt,
    fill=white,
    font=\bfseries,
    inner xsep=6pt,
    inner ysep=2pt
  ] at ([xshift=-0.2\linewidth]box.north) {UHF};

\end{tikzpicture}

\paragraph{Message authentication codes.} Similar to the notion of a universal hash function is that of a message authentication code (MAC). The origin of these objects is somewhat nebulous, but appears to have first entered the mainstream in various (now withdrawn) standards in the mid-to-late 1980s. Formally, a MAC is a trio of algorithms $(\kg,\macsign,\macvfy)$ that work as follows: $\kg$ generates a private key (which is no longer explicitly required to be sampled uniformly at random) to be distributed to the honest parties; $\macsign$ takes as input a key $k$ and a message $m$, and returns a tag $t$; and $\macvfy$ takes as input a key $k$ and a message-tag pair $(m,t)$, and outputs whether it considers this pair valid or not. The $\macvfy$, similarly to UHFs, operates according to what has been called\cite{crypto-book} \textit{canonical verification}; that is, in order to verify whether a message-tag pair $(m,t)$ is valid, check whether $\macsign(k,m)=t$ or not. The obvious notion of correctness should hold, that is, for each key $k$, we should have that for any message $m$, $\macvfy(k,m,\macsign(k,m))=\top$. 

The most common security notion is called $\mf{euf-cma}$ security, which stands for `existential unforgeability in the presence of a chosen message attack'. To decipher this rather opaque phraseology, note that `unforgeability' refers to the difficulty of an attacker not in possession of the secret key in producing a valid message-tag pair. The term `existential' is borrowed from a notion introduced in the context of digital signature\cite{goldwasser-euf-cma}, and describes the notion that a valid forgery for \textit{any} message will do -- even if that message is somehow irrelevant to the application (``random or nonsensical'', in the words of the authors who introduce this notion). Finally, a chosen message attack (more properly an \textit{adaptive} chosen message attack) references the notion that an attacker should be able to see valid tags on messages of its choice. Bellare and Rogaway justify enforcing such an apparently strong notion quite early in the history of formalising cryptography\cite{br-modern-cryptography}, and also advocate for the separate treatment of `replay' attacks; that is, a `forgery' that consists of the attacker relaying an existing, honestly-generated message-tag pair to the receiver.

The standard security game then goes like this. A challenger runs the $\kg$ algorithm, and receives queries (we may keep track of how many) on messages of the attacker's choice. For each of these messages, the challenger generates tags according to the key it has generated and returns them to the attacker. Eventually, the attacker generates a challenge message-tag pair $(m^*,t^*)$; if this pair validates under the key the challenger generated, and $m^*$ has not previously been queried to the challenger the game is won. Otherwise, it is lost. We can then characterise the security of a MAC by the advantage of polynomial-time attackers in such a game.  

The standard game-based security notions often suffice for composable security in some precisely defined sense. Interestingly, this is not the case here -- notice that when the attacker submits its challenge pair, it implicitly learns if that pair was valid by the outcome of the game; in other words, it has access to a one-time verification oracle. It turns out\cite{composable-review} that we require as many verification queries as tagging queries for composable security.

Unlike with UHFs, MACs offer computational, rather than unconditional security. However, this is the one-way function variant of computational security, and so might be regarded as a relatively modest assumption. The trade-off is that the security guarantees permit an essentially unbounded number of uses of each key -- or at least there is no hard limit on the number of uses, as with UHFs. There is also no hard limit on the size of the keys with respect to the messages, and indeed the designer of a MAC is free to boast as she likes of her scheme's short private keys, provided she can provide a compelling argument that the relevant computational assumption is sound. 

There are a large number of $\mac$s in the literature constructed from a wide variety of purported one-way functions. Important examples include those based on hash functions\cite{FIPS198-1}, and those based on block ciphers\cite{omac, cbc-mac}. 

\begin{tikzpicture}
  \node[
    draw,
    rounded corners=6pt,
    fill=yellow!20,
    inner sep=8pt,
    text width=0.75\columnwidth,
    align=left
  ] (box) {

    \vspace{6pt}

    \textbf{security} \dotfill\ computational (OWFs)\\
    \textbf{set-up} \dotfill\ PSKs\\
    \textbf{composability} \dotfill\ known\\
    \textbf{public verifiability} \dotfill\ no\\
    \textbf{scalability} \dotfill\ poor
  };

  \node[
    draw,
    rounded corners=3pt,
    fill=white,
    font=\bfseries,
    inner xsep=6pt,
    inner ysep=2pt
  ] at ([xshift=-0.2\linewidth]box.north) {MAC};

\end{tikzpicture}

\paragraph{Authenticated Encryption.}
Finally, we discuss the functionality of authenticated encryption, where authentication \textit{and} privacy are achieved from symmetric keys. It is a classical result\cite{crypto-book} that the preferred means of achieving this is by first encrypting, and then computing a tag on the resulting ciphertext. One could accomplish unconditional security by using the one-time pad for encryption, and a universal hash function to compute the tag, but the relevant PSKs would be one-use only. Alternatively, one could re-use PSKs at the cost of achieving only computational security.  

Composable security is hinted at in various introductions to CC\cite{portmann_renner_qkd_proof}, where it is shown that the one-time pad protocol achieves privacy in the presence of an active adversary, given an authenticated channel. Indeed, the CC framework is a rather native framework in which to conduct such an analysis, and composability has indeed been demonstrated\cite{composable-review}.\medskip

\begin{tikzpicture}
  \node[
    draw,
    rounded corners=6pt,
    fill=yellow!20,
    inner sep=8pt,
    text width=0.75\columnwidth,
    align=left
  ] (box) {

    \vspace{6pt}

    \textbf{security} \dotfill\ {\small unconditional/computational}\\
    \textbf{set-up} \dotfill\ one-use PSKs/PSKs\\
    \textbf{composability} \dotfill\ known\\
    \textbf{public verifiability} \dotfill\ no\\
    \textbf{scalability} \dotfill\ very poor/poor
  };

  \node[
    draw,
    rounded corners=3pt,
    fill=white,
    font=\bfseries,
    inner xsep=6pt,
    inner ysep=2pt
  ] at ([xshift=-0.2\linewidth]box.north) {AE};

\end{tikzpicture}

\subsection{Public-key methods} \label{sec:public}
We now consider methods of authentication based on public-key cryptography. This time, there is a public and private key -- as discussed, we need to assume that the communication parties agree which public key they are communicating with respect to.

\paragraph{Digital signatures.} First and most importantly in this regime are \textit{digital signature schemes} (DSSs). Without these remarkable objects, modern authenticated communication at scale would not be possible.

The concept of a DSS was introduced by Diffie and Hellman\cite{new-directions}, and the first instantiation was introduced as part of the RSA suite of cryptosystems\cite{rsa-signatures}. They work as follows: a key generation algorithm produces a key pair $(pk,sk)$, called the \textit{public key} and \textit{private key}. As the names suggest, the public key can be shared arbitrarily, but the private key must be kept secret. Unlike in the symmetric-key case, the algorithm generating tags and the verification algorithm are distinct: to \textit{sign} a message, we call on a signing function that takes as input the message to be signed and the secret key, and output a tag called a \textit{signature}. The verification algorithm takes as input a message-signature pair and the public key. This is why faithful distribution of the public key is required -- if a public key is not bound to an identity then an attacker can insert their own choice of public key, for which they know the corresponding secret key, and so forgeries are trivial. The ability to share the public key arbitrarily is what gives the scheme its public verifiability.

Again, we insist on $\mf{euf-cma}$ security\cite{goldwasser-euf-cma}. Notice that this time we trivially get verification queries, as a consequence of the public verifiability. Indeed, for this reason, showing that composable security holds with respect to the standard security notions is somewhat simpler than with MACs, and indeed composability has been demonstrated in various frameworks and settings\cite{canetti-composable-signature,maurer-unilateral,cc-signatures}.

The public verifiability of DSSs allows us to construct a PKI in the following way. Suppose $n$ parties have `registered' with a trusted authority, in the following sense: for party $i$, the trusted authority receives an authentic copy of that party's public key, and the party receives both the authority's public key, and the authority's signature on the message ``\texttt{party $i$'s public key is $pk_i$}''. This message, together with the authority's signature, is called a \textit{certificate}. Parties can now communicate authentically using only insecure channels by transmitting their certificates along with any message-tag pairs; because registered parties have a copy of the authority's public key, and the signature is publicly verifiable, the receiver can check that they are verifying the message-tag pair with respect to the correct public key. Moreover, this registration is one-time, regardless of the size of the existing network, and allows authenticated communication over insecure channels to any other registered party. Indeed, certificate authorities need not restrict themselves to attesting authenticity of end user's public keys -- they can also produce certificates with respect to other certificate authority's public keys (so-called \textit{intermediary} certificates). In this way, users can still establish authentic communication over insecure channels even when they are not registered with the same certificate authority, provided these intermediary certificate authorities are registered with each other. 

Probably the two most important DSSs are RSA signatures\cite{rsa-signatures}, whose security is based on the difficulty of integer factorisation; and ML-DSA\cite{fips204} (originally Crystals-Dilithium), which is one of two DSS standards recently published by NIST, effectively representing the first outcome of the long-running PQC standardisation process. Of the two extant standards, ML-DSA is the most efficient and most suitable for general purpose applications. However, it remains troublingly space-inefficient, with its public keys and signatures having length about 1.3kB and 2.4kB, respectively -- compare this to the roughly 256B required for RSA signatures. The crucial advantage of ML-DSA is that it is based on the presumed difficulty of certain problems in high-dimensional lattices, which we do not currently believe to be vulnerable to quantum cryptanalysis. This is in stark contrast to integer factorisation, which is famously\cite{shors-algo} quantum-vulnerable for a sufficiently large quantum computer.

\begin{tcolorbox}[colback=blue!10,colframe=blue!50, title=ML-DSA]
    In what follows, $Z_q$ is the ring $\ZZ_q[X]/\langle X^n+1\rangle$, and $\mathcal{Q}$ is a finite-field variant of the function that returns a quotient, acting component-wise on polynomial coefficients and entries of vectors and matrices. 
    \begin{itemize}
        \item \textbf{Key generation}. Sample a random matrix $A\getsr Z_q^{k\times l}$, and short vectors $\textbf{s}_1\getsr Z_q^l,\textbf{s}_2\getsr Z_q^k$. Compute $\textbf{t}\gets A\textbf{s}_1+\textbf{s}_2$. The public key is $(A,\textbf{t})$, the private key is $(\textbf{s}_1,\textbf{s}_2)$.
        \item \textbf{Signing $m$.} Sample a random short vector $y\in Z_q^l$ and compute $\textbf{w}_1\gets\mathcal{Q}(A\textbf{y})$, and set $c\gets H(\textbf{w}_1||m)$ (for a suitably defined public hash function $H$). If $\textbf{z}\gets\textbf{y}+c\textbf{s}_1$ is too long, or the vector $A\textbf{y}-c\textbf{s}_2$ is too long, abort and sample a different $\textbf{y}$. The signature is $\sigma=(\textbf{z},c)$.
        \item \textbf{Verifying $(m,\sigma)$.} Given $m$ and $(\textbf{z},c)$, compute $\textbf{w}'_1\gets\mathcal{Q}(A\textbf{z}-c\textbf{t})$, and accept if $c=H(m||\textbf{w}'_1)$.
    \end{itemize}
        Correctness follows from the rejection sampling in the signing process; the vectors are not too long, so is there no `wraparound' in terms of the quotient.
\end{tcolorbox}

Interestingly, it is possible to construct digital signatures from one-way functions, via the method of Lamport\cite{lamport-signatures}. Suppose $f:Y\to Z$ is a one way function, and the message is a single bit $b$. The signer samples $y_0,y_1$ at random and computes $z_i=f(y_i)$: they keep $y_0,y_1$ as their private key and broadcast $z_0,z_1$ as the public key. In order to sign a bit $b$, they set the signature as $\sigma=y_b$ and send $b,\sigma$. The verifier checks that $f(\sigma)=z_b$.

Note that the signature \textit{is} the private key, so in the textbook version these signatures are one-shot. Moreover, the public key is a function of the size of the message, because a value from the output space is needed for each bit of the message, and so parameters are generally huge. Nevertheless, it is desirable to have digital signatures from weaker assumptions, especially since the consensus view is that the existence of classical one-way functions will not be disproven by quantum computation. Various techniques exist in the literature to make these objects multi-shot and reduce their somewhat inevitable computational load; the most important example is the NIST standard SLH-DSA\cite{fips205}. 

Finally, we note that there has been increasing interest in versions of digital signatures that hope to securely provide advanced, essentially multi-party functionalities, both in the pre and post-quantum regime. In the spirit of this review we highlight examples of composable security analysis on these lines\cite{composable-threshold-signatures-1,composable-threshold-signatures-3, composable-ring-signatures-1,composable-ring-signatures-2, composable-blind-signatures-1,composable-blind-signatures-2}. \medskip

\begin{tikzpicture}
  \node[
    draw,
    rounded corners=6pt,
    fill=yellow!20,
    inner sep=8pt,
    text width=0.75\columnwidth,
    align=left
  ] (box) {

    \vspace{6pt}

    \textbf{security} \dotfill\ computational (both)\\
    \textbf{set-up} \dotfill\ one-time authentication\\
    \textbf{composability} \dotfill\ known\\
    \textbf{public verifiability} \dotfill\ yes\\
    \textbf{scalability} \dotfill\ good
  };

  \node[
    draw,
    rounded corners=3pt,
    fill=white,
    font=\bfseries,
    inner xsep=6pt,
    inner ysep=2pt
  ] at ([xshift=-0.2\linewidth]box.north) {DSS};

\end{tikzpicture}

\paragraph{KEM-then-MAC.} Notice that in the case of a digital signature, the sender possesses the private key. If we want to study privacy in the public-key regime this changes\cite{rsa-signatures}: the sender must have an authentic copy of the public key distributed to them, with which they can encrypt; the receiver keeps the associated private key, with which they can decrypt. Owing to the cost of public-key cryptography, we do not typically encrypt messages directly, instead using these privacy-preserving public-key mechanisms to establish a secret key for use with symmetric key cryptography. There are two important modes of operation in this vein: a non-interactive key exchange (NIKE)\cite{nike}, of which Diffie-Hellman key exchange\cite{new-directions} is the principal exemplar; and a key encapsulation mechanism (KEM), which we define below. In a pre-quantum world, the former is the standard; owing to a lack of post-quantum alternatives (with some exceptions\cite{csidh,swoosh}), we focus on KEMs in a post-quantum world. The salient example is NIST's post-quantum standard, ML-KEM\cite{fips203}.

A KEM, then, is a trio of algorithms $(\kg,\kemenc,\kemdec)$. $\kg$ produces a key pair $(pk,sk)$; by convention, the \textit{encapsulation} algorithm $\kemenc$ takes as input only the public key $pk$, and produces a random value and a ciphertext\footnote{In almost all cases this is the result of generating a random value and calling a public key encryption algorithm taking the key and that random value as input. Nevertheless, the convention is to write the encrypted value as distinct to the process of encapsulating.}; the \textit{decapsulation} algorithm $\kemdec$ takes as input the secret key and a ciphertext, and outputs the corresponding random value associated to that ciphertext. Writing ciphertexts as $z$ and random values as $\mu$, it should hold that for a public key pair $(pk,sk)$ and $(z,\mu)=\kemenc(pk)$, we have $\kemdec(sk,z)=\mu$.

For both NIKEs and KEMs, the apparently privacy-preserving mechanism can be used to bootstrap authentication according to the following intuition: if a sender possesses the receiver's authentic public key, they can implicitly be sure that only the receiver can read the encrypted symmetric key, and so any communication secured with that symmetric key must have come from the receiver. This is quite an old idea, and has been examined both in terms of NIKEs\cite{optls,static-dh}, and with KEMs in\cite{ake-with-pq-kems,ake-with-kems,pq-wireguard}. The idea of using MACs with secret keys derived from an authenticated KEM exchange appears to originate in\cite{bellare1998modular}.

An obvious question is why one would do this. After all, this method faces a disadvantage in terms of round trips: we have to send a public key, wait to receive a ciphertext, and then begin authentic communication, whereas in the signature regime we can send a public key and authenticated data in the same flight. The answer is that, in contrast to their pre-quantum equivalents, the NIST standard KEMs are much more space-efficient than their signature counterparts. For example, when we consider the objects that are required to be sent on-the-wire, at the lowest security level ML-KEM has a public key of length 800B, and a ciphertext of length 768B; ML-DSA has public keys of length 1.3kB and signatures of length 2.4kB. Meanwhile, the hash-based signature standard can yield signatures on the order of tens of kilobytes, depending on the underlying hash function. 

The KEM-then-MAC approach is therefore suitable in an environment when post-quantum security and bandwidth are a concern; for example, the single-server-many-client scenario AKE setting was studied in\cite{kem-tls}, and again from a composable perspective in\cite{ckake}. The techniques in this latter work were developed concurrently with the composable treatment of KEM-then-MAC in our companion work\cite{composable-review}.

\subsection{Unconditional and transferable authentication.}\label{sec:qdss}
There has been some work on using quantum information to authenticate classical data, notably in the sense of ``quantum digital signatures'' \cite{qdss}, as originally proposed by Gottesman. These are not the same thing as a scheme that produces secure, publicly verifiable tags on quantum data, which are known to be impossible to construct\cite{q-signcryption}. Instead, they can be seen as a quantum generalisation of digital signatures derived from one-way functions, where the difficulty of inverting the analogous one-way function is an unconditional guarantee from Holevo's theorem, as opposed to an assumption on the existence of one-way functions.

The protocol requires verifying parties to receive \textit{a priori} quantum states to act as `public keys'; verification occurs as a function of this quantum public key. In the same way that, in the classical setting, one can only hope for security up to the validity of public keys, the correct quantum state must arrive with the verifier -- in other words, the protocol assumes quantum authentication in its set-up. For unconditional security guarantees to be preserved, this means pre-shared keys need to be distributed before a protocol run. Note also that this means that the verifier needs long-term quantum memory, which puts this protocol currently out of reach.

One crucial property provided by this protocol is a limited degree of public verifiability, in the sense that a signature recipient is able to convince a limited number of other parties that a signature it receives is valid, provided they also possess the appropriate quantum public key (we might call this limited public verifiability \textit{transferable} verifiability). The use case for this protocol, then, is a scenario in which one requires both unconditional security \textit{and} a degree of public verifiability; a combination of properties that we have not yet seen provided. 

The requirement of long-term, high-quality memory is infeasible, and effectively renders the early attempts\cite{qdss} unusable. Various improvements were later made: removing the quantum memory requirement\cite{qdss-improv-1}, demonstrating that a comparable scheme can be realised only on QKD (i.e., extant) hardware\cite{qdss-improv-2}, and removing the requirement for quantum authentication\cite{qdss-improv-3}. For unconditional security purposes, none of these protocols succeed in dispensing with requiring pre-shared keys between all participants (as we might expect). In the spirit of this review, we might ask: do there exist purely classical schemes with comparable properties, and do we lose anything by prohibiting the use of quantum technologies?

The answers to these questions appear to be: yes, and no. There do exist unconditionally secure authentication schemes providing transferable verifiability, which are called ``unconditionally secure signatures''\cite{uss-1,uss-3,uss-4,uss-5}. With similar (rather stringent) pre-shared key set-up requirements, these schemes achieve unconditional security and limited public verifiability to quantum digital signatures. Moreover, there does not appear to be a similar operational advantage to introducing quantum technology as in the analogous QKD case (that is, there is no phenomenon analogous to quantum key expansion). As such we recommend the most modern\cite{uss-5}, fully classical USS for this somewhat niche use-case. \medskip

\begin{tikzpicture}
  \node[
    draw,
    rounded corners=6pt,
    fill=yellow!20,
    inner sep=8pt,
    text width=0.75\columnwidth,
    align=left
  ] (box) {

    \vspace{6pt}

    \textbf{security} \dotfill\ unconditional\\
    \textbf{set-up} \dotfill\ PSKs\\
    \textbf{composability} \dotfill\ unknown\\
    \textbf{public verifiability} \dotfill\ limited\\
    \textbf{scalability} \dotfill\ poor
  };

  \node[
    draw,
    rounded corners=3pt,
    fill=white,
    font=\bfseries,
    inner xsep=6pt,
    inner ysep=2pt
  ] at ([xshift=-0.2\linewidth]box.north) {USS};

\end{tikzpicture}

We close this discussion by noting that the composable security of these schemes remains an open question.

\subsection{On Identity}\label{sec:on-identity} We have exhibited various cryptographic mechanisms in the message-tag paradigm, which prevent a message being altered in transit. However, if the wrong person has access to the set-up of this mechanism, forgeries are trivial: for example, if an honest party shares a pre-shared key with an adversary then forgeries are trivial. Similarly, if an honest party convinced that the wrong public key is bound to a particular identity, forgeries are trivial, just by running the honest tagging algorithms. In other words, these mechanisms do not provide assurance of the identity of one's communication partner by themselves -- instead, identities are implicit in the set-up resources.

\section{Authentication of quantum messages} \label{sec:quantum-message}

So far, we have discussed the authentication of classical data, where a classical message is accepted only if it was generated by the legitimate sender. In a quantum network, however, the transmitted object may itself be a quantum state, and one may want to check the authenticity of a quantum message\footnote{The range of applications of such a functionality is ostensibly much sparser than the widespread requirement for classical authentication; see Section~\ref{sec:qma-apps}}. As with any quantum functionality, authentication of a quantum message comes with fundamental challenges and differences from its classical counterpart. For instance, the receiver cannot simply read the state, and verification procedures will often disturb the state itself. Having a quantum message also changes things from the perspective of the adversary, since the quantum message is now often not \emph{public} information. Hence, for a quantum message, the desired authentication functionality seems to be that Alice inputs an unknown quantum state $\rho$, and Bob either rejects or accepts and obtains a state that is essentially the same as $\rho$. However, against quantum adversaries, this should also include any entanglement that $\rho$ may have had with an external reference system.

The first formal treatment of this problem is due to Barnum, Cr\'epeau, Gottesman, Smith and Tapp~\cite{barnum2002authentication}. The quantum authentication (QA) schemes defined here, and in most of the works that followed, are symmetric-key schemes. As such, a quantum authentication scheme consists of keyed encoding and verification procedures. Alice uses a shared classical secret key to encode an $m$-qubit message into a larger transmitted system, and Bob uses the same key to decode and either accept or reject. Completeness requires that, without interference, Bob accepts and recovers the original state. Soundness requires that, for any adversarial quantum operation on the transmitted system, Bob's final state is close to the ideal behaviour: either Bob rejects, or the accepted message is close to the state Alice sent. Their original construction can be understood as the quantum analogue of \emph{encrypt-then-authenticate}; however, there is an important reason for this. They show that to authenticate a quantum message, encryption is necessary. This already illustrates a sharp difference from classical authentication. Classically, authentication and secrecy are independent: a Wegman--Carter tag or a MAC can authenticate a plaintext message while leaving the message publicly readable. Quantumly, this separation collapses. Barnum et al.~provide intuition that any quantum authentication scheme with small error must also be a good quantum encryption scheme~\cite{barnum2002authentication}. The intuition is simple but powerful. If an adversary can distinguish the authenticated encodings of two states, say $\ket{0}$ and $\ket{1}$, then she can coherently condition an operation on this information and change a superposition such as $\ket{0}+\ket{1}$ into $\ket{0}-\ket{1}$ without being detected. In other words, information leakage can be converted into undetectable coherent tampering.

The construction they propose combines the quantum one-time pad (QOTP) and error-correcting codes. Alice applies a quantum one-time pad and then encodes the encrypted state into a randomly chosen stabilizer error-detecting code, together with a hidden syndrome. Bob reverses the encoding, checks the syndrome, and only then decrypts. The stabilizer code is not being used primarily to correct noise, but to detect adversarial tampering. This motivates the language of \textit{purity-testing codes}: the code family is chosen so that any non-trivial Pauli error is detected for all but a small fraction of keys.

This has two immediate consequences. First, authenticating $m$ qubits requires essentially $2m$ bits of secret key, matching the cost of the quantum one-time pad up to lower-order authentication overheads. Second, it rules out a natural quantum analogue of digital signatures for unknown quantum states. A classical digital signature works because verification can be public: anyone can read the message and check the signature, but only the signer can produce a new valid signature. For unknown quantum messages, this is impossible in the usual sense. If a recipient can efficiently verify and extract the signed quantum state, then they can replace it by another state and run the extraction circuit backwards, producing a valid signed state for a different message. This rules out transferable, publicly verifiable signatures on arbitrary unknown quantum states~\cite{barnum2002authentication}.

\paragraph{Quantum non-malleability as the neighbouring functionality.}
Quantum authentication is closely related to \textit{non-malleable encryption}. Non-malleability is an additional property of encryption schemes that requires the adversary to be unable to manipulate the ciphertext in a way that would make predictable or useful changes to the decoded plaintext. Not all encryption schemes are non-malleable, either quantumly or classically. A good quantum example is the quantum one-time pad (QOTP)~\cite{qotp-Mosca2000}, which provides information-theoretic perfect secrecy, but if the adversary applies a Pauli operator to the ciphertext, the corresponding Pauli operator is applied to the plaintext after decryption. Thus, the adversary learns nothing, but can still implement a controlled transformation, which makes QOTP malleable.

Ambainis, Bouda and Winter introduced non-malleable encryption of quantum information to rule out precisely this behaviour~\cite{ambainis2009nonmalleable}. Informally, non-malleability asks that any effective operation induced on the plaintext by attacking the ciphertext is essentially trivial: either the adversary does nothing, or the adversary replaces the plaintext with a fixed state independent of the original message. In the case where the encryption scheme is a unitary transformation, they prove an elegant characterization, showing that perfect non-malleable quantum encryption is achieved by encryption with a unitary $2$-design, and that the $2$-design property is sufficient for achieving this functionality. In comparison, the QOTP is a unitary $1$-design, which is sufficient for secrecy but not for non-malleability. In terms of key-size requirements, secrecy requires roughly $2\log d$ bits for a $d$-dimensional quantum system, whereas unitary non-malleability requires roughly $4\log d$ bits.

Alagic and Majenz later introduced an extended notion of quantum non-malleability~\cite{alagic2017quantum}, which is strictly stronger than the original notion~\cite{ambainis2009nonmalleable}. They argue that for general, non-unitary schemes, the definition of Ambainis et al.\ can allow a \textit{plaintext injection} attack, in which the adversary forces the receiver to decrypt to a plaintext of the adversary's choice. In contrast, their definition allows adversarial side information, and the security is given against these attacks as well. However, despite the stronger definition, the unitary $2$-design scheme also satisfies the Alagic--Majenz definition of non-malleability. Another interesting result is shown: in the quantum setting, unlike in the classical setting, non-malleability itself implies secrecy.

The relationship between non-malleability and extended versions of quantum authentication was discussed by Ambainis et al~\cite{ambainis2009nonmalleable}, and clarified further by Alagic and Majenz~\cite{alagic2017quantum}. In order to unpick these relationships we first remark that many mentions of \emph{quantum authentication} in the literature seem to refer to non-malleability-based quantum authentication schemes built from the non-malleable encryption introduced in the work of Ambainis et al., rather than to the original terminology of Barnum et al~\cite{barnum2002authentication}. Notably, these two works show that quantum authentication can be built from non-malleable encryption by adding a tag register before applying non-malleable encryption. Thus, the main distinction between non-malleability and authentication can be seen in the rejection mechanism. Non-malleable encryption says that attacks cannot induce a useful transformation of the plaintext. Authentication says, more operationally, that attacks should either leave the plaintext intact or be detected.

There are also two other main quantum authentication notions that are stronger than the original definition~\cite{barnum2002authentication}. The definition associated with Dupuis, Nielsen and Salvail~\cite{dupuis2012actively}, often called DNS authentication, is simulation-based and asks that, on average over the secret key, any adversarial attack is equivalent to an ideal attack that does not touch the authenticated plaintext. In their work, they mainly use quantum authentication as a verification subroutine for a secure two-party quantum computation protocol. If all authentication checks pass, the authenticated values collapse to the correct values; otherwise, the protocol aborts. The authentication code must also allow computation through the authentication and checking without revealing the authenticated data, which is why Clifford-based authentication codes play a central role.

\begin{figure}[H]
    \centering
    \begin{tikzpicture}[
        node distance=1.8cm and 1.8cm,
        every node/.style={font=\small},
        box/.style={
            draw,
            rectangle,
            rounded corners=3mm,
            fill=red!20,
            minimum height=9mm,
            align=center,
            inner sep=3pt
        }
    ]
    
    \node[box, minimum width=3.0cm] (ta)
        {Total\\Authentication};
    
    \node[box, minimum width=3cm, below=1.6cm of ta] (dns)
        {DNS};
    
    \node[box, minimum width=3cm, below=1.6cm of dns] (qa)
        {QA};
    
    \node[box, minimum width=3.0cm, right=2cm of ta] (am)
        {QNME (AM)};
    
    \node[box, minimum width=3.0cm, right=2cm of dns] (abw)
        {QNME (ABW)};

    \draw[<-, dashed] (ta.east) -- (am.west) node[midway, above] {+ tag};
    
    \draw[->] (ta) -- (dns);
    
    \draw[->,dashed] (abw) -- (dns) node[midway,above] {+ tag};
    
    \draw[->] (am.250) -- (abw.110);
    \draw[->, crossed] (abw.70) -- (am.290);
    
    \draw[->] (dns.250) -- (qa.110);
    \draw[->, crossed] (qa.70) -- (dns.290);
    \end{tikzpicture}
    \caption{The known relations between notions of quantum authentication and the related notion of quantum non-malleable encryption. A dashed line between two notions and a ``+ tag'' label indicate that an authentication scheme can be constructed from the former QNME scheme by adding a tag register\cite{ambainis2009nonmalleable,alagic2017quantum}. A non-dashed, non-crossed arrow between two notions means that the former implies the latter; a non-dashed, crossed arrow between two notions means that the former is strictly weaker than the latter.}
    \label{fig:qa-notions}
\end{figure}
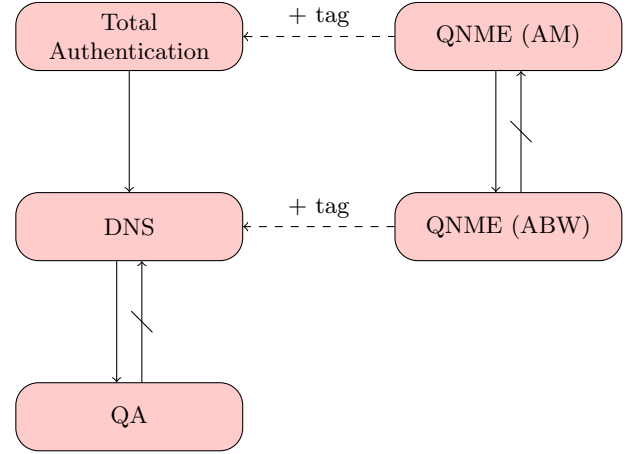

Garg, Yuen and Zhandry~\cite{garg2016new} introduced stronger definitions in the spirit of the stronger non-malleability notion discussed by Alagic and Majenz~\cite{alagic2017quantum}. Their strongest notion, \textit{total authentication}, says that any real adversary can be reduced to an oblivious adversary that does not act on the authenticated message at all and only acts on its own side information. One important property of this scheme is key reusability: under total authentication, if Bob accepts, the entire secret key can be safely recycled. Their original construction used approximate unitary $8$-designs and tag qubits. Alagic and Majenz later showed that $2$-designs with tags already suffice for total authentication, and also imply the weaker Dupuis--Nielsen--Salvail-style authentication notion after a mild adaptation of the rejection procedure~\cite{alagic2017quantum}. We detail the relationship between these notions in Figure~\ref{fig:qa-notions}.

\subsection{$2$-design schemes}

Let us now discuss the simplest construction for quantum authentication, which is based on unitary $2$-designs. Alice appends $s$ qubits in state $\ket{0^s}$, applies a secret unitary from a $2$-design to the message-plus-tag system, and sends the result. Bob applies the inverse unitary and measures the tag; if the tag is still $\ket{0^s}$, he accepts and outputs the message, otherwise he rejects. Alagic and Majenz show that this ``$2$-design plus tag'' construction yields quantum authentication in the strongest adversarial model discussed earlier~\cite{alagic2017quantum}. We thus introduce this $2$-design scheme as our candidate construction for the functionality of quantum authentication.

\begin{tcolorbox}[colback=blue!10,colframe=blue!50, title=$2$-design authentication with tags] \label{box:twodesign-auth}
    Let $M$ be the message register and let $T$ be an $s$-qubit tag register. Let $\mc{D}=\{U_k\}_{k\in\mc{K}}$ be an exact or approximate unitary $2$-design on $M\otimes T$, indexed by a shared classical secret key $k$.

    \begin{itemize}
        \item \textbf{Authentication.} To authenticate a quantum message $\rho_M$, Alice prepares
        \[
            \rho_M \otimes \ket{0^s}\!\bra{0^s}_T,
        \]
        applies $U_k$, and sends
        \[
            \sigma_{MT}=U_k\big(\rho_M \otimes \ket{0^s}\!\bra{0^s}_T\big)U_k^\dagger
        \]
        through the insecure quantum channel.

        \item \textbf{Verification.} Bob applies $U_k^\dagger$ to the received register and measures the tag register $T$ in the computational basis. If the outcome is $0^s$, he accepts and outputs the remaining message register $M$. Otherwise, he rejects and outputs $\bot$.
    \end{itemize}
\end{tcolorbox}

This scheme simultaneously explains non-malleable quantum encryption, quantum authentication, and the route to key recycling under stronger definitions. It is not necessarily the most key-efficient construction in every regime: the original purity-testing-code construction of Barnum et al.\ is asymptotically optimal in the leading $2m$ key term for authenticating $m$ qubits~\cite{barnum2002authentication}. The advantage of the $2$-design scheme is conceptual and modular: it makes the relationship between secrecy, non-malleability, authentication and key recycling especially transparent. It is also worth mentioning that although, strictly speaking, this scheme has not been proven to achieve composable security, the simulation-based nature of the definitions of~\cite{dupuis2012actively,alagic2017quantum} suggests that this scheme plausibly satisfies the desired properties in a composable setting.

In network terms, quantum message authentication should therefore be viewed as a resource-intensive but rich primitive. It is the right functionality whenever quantum data itself needs to be validated, for example in secure multi-party schemes. It is not a scalable replacement for classical authentication in QKD control traffic, and it does not provide a quantum analogue of public digital signatures.

\begin{tikzpicture}
  \node[
    draw,
    rounded corners=6pt,
    fill=yellow!20,
    inner sep=8pt,
    text width=0.75\columnwidth,
    align=left
  ] (box) {

    \vspace{6pt}

    \textbf{security} \dotfill\ unconditional\\
    \textbf{set-up} \dotfill\ {\small pre-shared classical key}\\
    \textbf{composability} \dotfill\ {\small plausible in some variants}\\
    \textbf{public verifiability} \dotfill\ no\\
    \textbf{scalability} \dotfill\ poor\\
    \textbf{candidate construction} \dotfill\ {\small $2$-design scheme}
  };

  \node[
    draw,
    rounded corners=3pt,
    fill=white,
    font=\bfseries,
    inner xsep=6pt,
    inner ysep=2pt
  ] at ([xshift=-0.2\linewidth]box.north) {QAuth};

\end{tikzpicture}

\subsection{Signcryption}
Barnum et al.'s intuition that no public verifiability can exist quantumly, rather surprisingly, did not receive a completely formal treatment for some time. Nevertheless, it is now known that any `reasonable' definition of a quantum digital signature cannot simultaneously satisfy notions of completeness \textit{and} soundness\cite{q-signcryption}; in other words, there is no public verifiability of quantum data.

We can, however, hope to alleviate our set-up costs in large networks, because of so-called `signcryption' mode of securing quantum data\cite{q-signcryption}. The idea is that one generates a classical private-key/public-key pair, encrypts against a public key, and signs this encryption with one's private key. Because the public keys to be distributed are classical we can rely on a PKI to distribute them, and so there are available notions of scalable authentication for quantum data. There are, however, not many constructions in this area, and this remains an under-studied area of research. 

\section{Entity Authentication} \label{sec:identification}

Let us now turn our attention to \textit{entity authentication}, a somewhat specialised variety of authentication. We will focus on a specific aspect of this area, which is of particular interest from a quantum technologies perspective; nevertheless, the field is sufficiently protean to warrant a degree of disambiguation.

At least as far as the literature is concerned, the distinction between the terms ``entity authentication'' and ``classical message authentication''\footnote{Note that the latter term is our terminology, and has been referred to variously in the literature as simply ``authentication'', ``data origin authentication'', and ``message integrity''.} is not completely clean. For example, the term ``entity authentication'' in the seminal work of Bellare and Rogaway on authenticated key exchange\cite{bellare1993entity} is described as follows:

\begin{quote}
\textit{`Entity authentication is the process by which an agent in a distributed system gains confidence in the identity of a communication partner. More often than not, the entity authentication process is coupled with the distribution of a ``session key'' which the partners can later use for message confidentiality, integrity, or whatever else.'}
\end{quote}

Despite an acknowledgement of the separation between key establishment and entity authentication, this intuition is at bottom describing the problem of binding symmetric keys to communication partners. In our treatment this is a problem solved by classical message authentication, which handles the exchange of messages in an adversary-controlled environment; the binding of identities to the set-up assumptions, for example with a PKI, is possibly most akin to what is meant by ``entity authentication'' in this context.

In contrast, the \textit{Handbook of Applied Cryptography}\cite{menezes1996handbook} makes the separation more explicit. For these authors, entity authentication protocols are 

\begin{quote}
\textit{`...techniques designed to allow one party (the \emph{verifier}) to gain assurances that the identity of another (the \emph{claimant}) is as declared, thereby preventing impersonation'  
}
\end{quote}

From this point of view, classical message authentication suffices for entity authentication, just by sending an arbitrary classical message (say, \texttt{``I am Alice''}). On the other hand, this is rather tautological, since (as discussed in Section~\ref{sec:on-identity}) we have to assume that set-up assumptions correctly bind authenticated classical channels to identities -- so in some sense the entity authentication has `already happened'. Here we are helped by further discussion from the handbook\cite{menezes1996handbook}:

\begin{quote}
\textit{`A major difference between entity authentication and message authentication (as provided by digital signatures or MACs) is that message authentication itself provides no timeliness guarantees with respect to when a message was created, whereas entity authentication involves corroboration of a claimant’s identity through actual communications with an
associated verifier during execution of the protocol itself (i.e., in real-time, while the verifying entity awaits)'}
\end{quote}

With this in mind we advocate for the following distinction between entity authentication and classical message authentication is as follows: entity authentication is an interactive protocol which convinces verifiers of claimants' identities, without binding any data to identities; classical message authentication is a non-interactive protocol that provides a non-time-specific proof that a particular string of classical data is bound to a particular identity. 

Having established the clean separation, we note that the two functionalities are nevertheless closely linked: one can often compile interactive entity authentication protocols into non-interactive message authentication protocols\footnote{A famous example is the Fiat-Shamir transform\cite{fiat-shamir}, which converts three-pass interactive protocols to digital signatures.}; and a message authentication protocol suffices to construct an identification protocol by responding to challenges from a verifier with that protocol. We defer detailed discussion of these interactions to separate work -- since our focus here is on protocols that remain secure against quantum adversaries while exploiting resources naturally available in quantum networks, we focus instead on hardware-based approaches.

In the hardware regime, authentication is tied to access to a physical module rather than only to knowledge of a cryptographic key~\cite{pappu02}. This is natural for entity authentication because the goal is precisely to verify the presence or control of a particular party or a device. Among the different possible hardware assumptions for this task, Physically Unclonable Functions (PUFs) are particularly natural candidates. A PUF is a physical device whose challenge--response behaviour is determined by intrinsic manufacturing randomness, so that access to the genuine object functions as a device-level credential rather than as a copyable bit string~\cite{pappu02,gassend2002silicon,herder2014pufs,gao2020physical}. In this sense, PUF-based authentication offers a different kind of set-up assumption: instead of distributing and protecting only a secret string, one enrols a physical object whose behaviour is difficult to reproduce. However, classical PUFs are not automatically secure cryptographic objects. In particular, even strong PUFs with large challenge--response spaces can be vulnerable to modelling attacks, including machine-learning attacks, in which an adversary learns enough challenge--response pairs (CRPs) to emulate the device~\cite{ruhrmair2010modeling,ruhrmair2013modeling,gao2020physical}. Quantum and hybrid PUF-based protocols are motivated by this gap between the appealing physical intuition of PUFs and the difficulty of proving strong security for practical protocols. In the past decade or so, \emph{quantum hardware security} has emerged as a subfield in which several works aim to exploit quantum communication and genuinely quantum features, such as no-cloning and non-orthogonal state discrimination, to extend the notion of physical unclonability into the quantum setting and design quantum-secure protocols~\cite{skoric2012quantumreadout,goorden2014quantumsecure,nikolopoulos2021remote,arapinis2021qpuf,doosti2021client,chakraborty23,goswami2025hybrid}. Here, we mostly focus on the works that aim to provide provably secure entity authentication against quantum adversaries with minimal hardware assumptions. This makes them especially relevant for quantum networks, where quantum communication is already in place and using these quantum authentication protocols may provide features that are unique to the quantum world, or avoid the full overhead of general-purpose post-quantum public-key infrastructure.

\paragraph{Generic structure of hardware-based protocols.}
Most hardware-based entity-authentication protocols follow the same high-level structure. In an \emph{enrolment} or \emph{set-up} phase, the verifier locally interrogates a physical module and stores a private database of CRPs, after which the module is assigned or transferred to the prover. In the \emph{identification} phase, the verifier samples fresh random challenges from this database (with respect to some distribution) and sends them over an insecure channel. A legitimate prover uses the physical module to generate the corresponding responses, while an adversary must attempt to emulate a valid response without having access to the module, possibly by using some limited database they gathered from the device during transition. Finally, in the \emph{verification} phase, the verifier checks the returned evidence, usually with a noise-tolerant acceptance threshold. Completeness means that an honest prover is accepted with high probability, while soundness or security means that an adversary is accepted only with negligible (for instance, exponentially small in the system size or number of rounds), or otherwise explicitly bounded, probability.

\subsection{Quantum-PUF protocols}
The fully quantum analogue of a PUF is a quantum PUF (qPUF), where the physical object is modelled as a quantum process that maps quantum challenges to quantum responses. The idea of using hardware-specific quantum features, and the unclonability of quantum states, as a defence mechanism against a quantum adversary already appeared in early proposals for quantum PUFs~\cite{skoric2012quantumreadout,nikolopoulos2021remote}. Arapinis, Delavar, Doosti and Kashefi gave the first systematic cryptographic treatment of qPUFs, formalising different levels of quantum unforgeability and showing both provable constructions and impossibility results~\cite{arapinis2021qpuf}. In particular, they show that, in the quantum world, unlike in the classical setting, unitary qPUFs can provide provable security against arbitrary polynomial-time quantum adversaries. This corresponds to satisfying a notion of quantum unforgeability known as \emph{universal unforgeability}\footnote{In the original paper, this notion is referred to as selective unforgeability. However, in later work~\cite{doosti2021unified}, these quantum notions of unforgeability were unified, and the corresponding level is called universal unforgeability; selective unforgeability corresponds to an intermediate security level.}, where the device is secure against an adversary trying to learn the outcome of the PUF, on average, for any random quantum challenge. The no-go result, on the other hand, states that there exists no quantum device satisfying unforgeability if the adversary themselves can select the challenge.

This framework was then used to design exponentially secure client-server identification protocols based on qPUFs~\cite{doosti2021client}. These protocols follow the generic structure discussed above: the verifier stores quantum challenge--response information generated by the qPUF. In practice, in many cases, the challenge states can be stored classically, but the response states need to be stored in quantum memory; otherwise, the verifier would need the computational capability to perform full tomography, which is often not available. Authentication is then carried out by testing whether the prover can reproduce the corresponding quantum response states. The two variants of the protocols discussed in that work trade off the resources required by the verifier and the prover. For example, if the verifier is a high-resource party, such as a quantum server, they can perform quantum equality tests. In the second variant, where the verifier is relatively low-resource and has no quantum computer, they can verify a quantum server by delegating the quantum-computation part of the verification process to the server. The security goal is that a quantum polynomial-time adversary, even after polynomially many interactions with the qPUF, cannot later impersonate the legitimate holder except with exponentially small probability. These protocols provide a rather strong security guarantee: the adversary can perform an arbitrary quantum channel, including coherent and adaptive attacks, and the protocols are efficient in terms of the size of the quantum systems and the number of rounds. However, these strong theoretical guarantees come at the price of requiring large quantum memories that need to survive for at least the duration of a full authentication.

Subsequent work refined this picture in several directions. One direction concerns cryptographic primitives and constructions. Kumar, Mezher and Kashefi proposed a more efficient qPUF construction based on unitary $t$-designs, replacing the unrealistic requirement of Haar-random unitaries with more structured designs that provide security against bounded-query adversaries~\cite{kumar2021efficient}. Another interesting construction is a variant called ``classical readout of quantum PUF", initially introduced in~\cite{phalak2021quantum} and expanded in~\cite{pirnay2022learning}. This is an attempt to retain an unpredictable quantum process while removing the requirement of quantum memory for authentication. The idea behind this type of qPUF is that there exists a known, and perhaps simple, quantum circuit, for instance, a single-qubit rotation followed by a Hadamard gate. The uniqueness of the device comes from the noisy implementation of this circuit on hardware, which is unique and unknown. In other words, there is a noise channel $\Lambda_{\mathrm{id}}$ that acts as the unique qPUF. At first, this might seem like a special case of a general qPUF. However, the main difference is in the construction of the response: here, the responses are obtained by classical post-processing of measurement outcomes, or equivalently, as expectation values of the post-processed state after the unknown channel. This allows the response to be classical while the underlying process is still quantum and expected to be hard to forge. However, Pirnay et al.~\cite{pirnay2022learning}, after properly modelling this type of qPUF using a particular type of query known as a statistical query (SQ), show that these constructions can be broken using simple ML-based modelling attacks. One might think that this is due to the simplicity of the circuit structure proposed in this construction, and that, in general, this type of qPUF could still provide good security. However, later work~\cite{wadhwa2025learning} shows that these attacks can be extended using cryptanalysis based on quantum learning theory. In particular, the authors show that quantum PUFs that operate through quantum statistical queries can be broken using quantum learning attacks in most relevant regimes, even when the underlying circuit is rather complicated. This result highlights the importance of quantum communication and of the verification procedure used in qPUF-based protocols.

Furthermore, qPUFs based on random measurement constructions have also been proposed~\cite{ghosh2024existential}, where they achieve a stronger notion of unforgeability. For more information about comparisons between these models, and between ideal models and more experimentally accessible qPUFs, we refer the reader to~\cite{galetsky2022comparison,farre2025review}. qPUFs as a hardware assumption and as an alternative route to quantum security have also been compared to other quantum assumptions, the most similar one being the notion of pseudorandom unitaries (PRUs). In~\cite{doosti2022pseudorandomness}, the authors show that PRUs can be constructed from a family of qPUFs, which gives the only existing construction of pseudorandom unitaries that does not itself rely on classical computational assumptions, such as quantum-secure one-way functions. On the other hand, PRUs can be used practically for authentication instead of full qPUFs, while providing a similar level of security.

Although qPUFs are powerful primitives for quantum cryptography more broadly, they sit at the high-resource end of the spectrum for the relatively simple entity-authentication functionality that we proposed here. Their implementation may require genuinely quantum hardware, storage of quantum datasets, and verification procedures for quantum states. This makes them conceptually clean but not necessarily the most practical route for near-term quantum networks. These limitations have motivated more implementation-oriented hardware-based protocols, which we group under the term \emph{hybrid} protocols. These are schemes that combine quantum communication or quantum information-theoretic effects with classical or semi-classical hardware modules.

\subsection{Hybrid quantum-classical protocols}
An implementation-oriented direction is to combine ordinary classical PUFs with quantum communication. The main idea is to use a weak or realistic classical PUF inside a hardware module that creates a quantum interface. This allows the protocol to exploit quantum communication in order to obtain provable security against quantum adversaries, as well as against classical and quantum machine-learning attacks. In this direction, it is important to be cautious of the fact that the underlying classical PUF is not assumed to be sufficiently secure on its own. The point of the protocol design is precisely to boost the security of this imperfect hardware by using quantum constructions and quantum communication.

The first variant of these hybrid PUFs was proposed in~\cite{chakraborty23}, and is known as the \emph{hybrid locked PUF} (HLPUF) construction. In a hybrid PUF, the output of a classical PUF is encoded into non-orthogonal quantum states, typically BB84 states, although other mutually unbiased basis states can also be used. Thus, the overall construction produces a sequence of quantum states in response to a classical challenge. The conjugate coding prevents an adversary from perfectly learning the response, and as a result, can considerably boost the security of an imperfect classical PUF. However, as shown in~\cite{chakraborty23}, this simple construction alone is not sufficient to provide security against an adaptive quantum adversary. To address this, a \emph{quantum lock} is introduced to prevent adaptive attacks in which the adversary repeatedly queries the device in order to build many copies of the same response state. The resulting construction is called a hybrid locked PUF (HLPUF).

We refer to the entity-authentication protocol built from this construction as a \emph{prepare-and-measure hybrid PUF} protocol. Operationally, the new device takes as input a classical challenge together with a quantum unlocking state. A legitimate prover, holding the genuine device, uses the quantum state to unlock the hardware and, at the same time, can authenticate the verifier's communication and return the appropriate quantum response to authenticate themselves. The verifier then checks the response against its private database. As such, the protocol proposed in~\cite{chakraborty23} is a double-authentication scheme, where a successful round implies that both parties are authenticated. This gives a provable quantum communication advantage even when the underlying classical PUF is only weakly secure, and it also enables CRP reuse in certain regimes.

A more recent class of hybrid authentication protocols asks whether the quantum layer can be entanglement-based rather than prepare-and-measure. We call this category \emph{entanglement-based hybrid protocols}. This direction is motivated by advanced quantum networks, where entanglement distribution may already be part of the network infrastructure, and in particular by the possibility of integration with entanglement-based variants of QKD. Entangled states provide a useful feature, namely local indistinguishability: subsystems of perfectly distinguishable global states can be maximally mixed, and hence convey no information about the encoding. Goswami, Doosti and Kashefi use this idea to introduce two entanglement-based hybrid authentication protocols~\cite{goswami2025hybrid}.

The first is an \emph{offline} protocol. Here, the prover and verifier receive pre-distributed Bell states during the set-up phase. During authentication, no quantum communication is required: the prover uses the local PUF response to choose a measurement on their half of the entangled state and sends only a classical outcome to the verifier. The verifier checks consistency using its own half of the entangled state and its CRP database. The advantage is that authentication can later be performed using only classical communication; the cost is the trusted pre-distribution and storage of entanglement.

The second is an \emph{online} protocol, which removes the need for pre-distributed entanglement by introducing a hybrid entangled PUF (HEPUF). In this case, the hardware takes a classical challenge and produces an entangled state chosen from a locally indistinguishable set. One subsystem is sent to the verifier, while the prover keeps and measures the other subsystem according to the PUF response. Security is then based on the fact that an adversary who intercepts only part of the entangled state, or is restricted to local operations, cannot reliably learn the hidden response information. This gives a hardware-based authentication protocol tailored to entanglement-enabled quantum networks, with one-way quantum communication from prover to verifier.

The hybrid protocols all achieve exponential soundness amplification over repeated rounds, but with different constants and resources. In the prepare-and-measure HLPUF protocol~\cite{chakraborty23}, the cheating probability is bounded by the forging probability of the underlying classical PUF, multiplied by an exponentially small quantum factor coming from the non-orthogonal locking and response states. The offline entanglement-based protocol~\cite{goswami2025hybrid} gives a particularly clean ideal bound of order $(1/2)^m$ for an $m$-bit response, but requires pre-distributed Bell pairs and quantum storage while using only classical communication during authentication. This ideal bound is extended to imperfect, and potentially adversarially tampered, Bell pairs. If each shared state has fidelity at least $1-\epsilon$ with the target Bell state, then the cheating probability is bounded by $2^{-m(1-\mu(\epsilon))}$, where $\mu(\epsilon)$ is a function of the security parameter. The online HEPUF protocol removes this pre-distributed-entanglement assumption and uses one-way quantum communication instead; its soundness again decreases exponentially with $m$, with a base determined by the implementation noise and distinguishability parameter, typically written as $\bigl(\frac{1}{2}+\delta\sqrt{(1+4\delta^2)/2}\bigr)^m$. Another important security feature of these protocols is that, although the overall security model is hardware-based and is often stated against quantum polynomial-time (QPT) adversaries, the adversary acting on the quantum network during the authentication phase can be information-theoretically unbounded. The polynomial-time restriction mainly enters because the model allows the adversary to access or query the hardware during the setup or transfer phase, and this access must be bounded to prevent complete learning of the device behaviour. Once the setup phase is over, the authentication-phase security is essentially information-theoretic, meaning that an adversary who only attacks the public quantum and classical communication cannot break the protocol except with the stated soundness probability, unless the protocol aborts.

Although the entanglement-based variants have interesting features, the prepare-and-measure hybrid protocol is arguably the simplest and most versatile option, both conceptually and from the perspective of experimental implementation. Moreover, the prepare-and-measure hybrid protocol has been shown to provide challenge-reusability~\cite{chakraborty23}, while the same property for the entanglement-based protocols has not been explicitly proven. We therefore choose it as our quantum-secure candidate protocol for entity authentication and describe it more precisely below. Yet we kept the specification as general as possible to also capture the essence of other variants. 

\noindent\begin{tcolorbox}[colback=blue!10,colframe=blue!50, title=HLPUF entity authentication, label=box:hybrid-puf-auth]
    Let $f$ be a classical PUF enrolled by the verifier, and let $\mathsf{Enc}$ be a quantum encoding of PUF responses into quantum states. In this particular instance, $\mathsf{Enc}$ encodes each pair of classical bits into a BB84 state. Let $y_i=f(x_i)$ be the full PUF response for a given challenge $x_i$, and write the response as a concatenation $y_i = y_i^1 || y_i^2$.
    \begin{itemize}
        \item \textbf{Set-up.} The verifier samples challenges and records the corresponding PUF responses $y_i=f(x_i)$, forming a private CRP database $\{(x_i,y_i)\}_{i=1}^K$. The device is then quantum-locked to form an HLPUF and is sent to the prover over an insecure channel.
        \item \textbf{Authentication.}
        \textit{Challenge.} In each authentication round, the verifier samples a fresh challenge $x_i$ and sends to the prover
        \begin{itemize}
            \item the classical challenge $x_i$;
            \item the verifier's quantum unlocking state $\rho_V = \mathsf{Enc}(y_i^1)$.
        \end{itemize}
        \textit{Hardware response.} A legitimate prover queries the physical module with $x_i$ and $\rho_V$. If the HLPUF accepts the unlocking state, it produces the prover's response state
        \[
            \rho_P = \mathsf{Enc}(y_i^2).
        \]
        The prover sends $\rho_P$ back to the verifier.
        \item \textbf{Verification.} The verifier checks the received response state against $\mathsf{Enc}(y_i^2)$ using the private CRP database, and accepts the round only if the prescribed quantum verification test passes. The overall authentication succeeds when sufficiently many rounds pass the test.
    \end{itemize}
\end{tcolorbox}

Regarding composability, the specific qPUF, HLPUF and HEPUF entity-authentication protocols discussed above have not, to the best of our knowledge, been analysed in a composable-security framework. This remains an important open question, especially if such protocols are to be used as subroutines inside larger quantum-network functionalities. We note, however, that composable treatments of PUF-based cryptography do exist in other settings, for example, for classical PUF-based key exchange and UC commitments~\cite{magri2022everlasting}.

Scalability is also only partially understood. Most existing quantum and hybrid PUF-authentication protocols are formulated in a two-party, client-server setting, although extensions towards richer network architectures are hinted at in~\cite{goswami2025hybrid}. Because verification relies on a private database, the most natural architecture is a star network: each client node possesses a unique PUF-based module, while a central verifier stores the corresponding CRP database and can authenticate each node on demand. This gives linear scaling in the number of enrolled devices and is reasonable for server-client networks, but it introduces a central verification bottleneck and does not directly provide scalable peer-to-peer authentication. Extending these hardware-based techniques to decentralised or multi-party quantum networks remains open.\medskip

\begin{tikzpicture}
  \node[
    draw,
    rounded corners=6pt,
    fill=yellow!20,
    inner sep=8pt,
    text width=0.75\columnwidth,
    align=left
  ] (box) {

    \vspace{6pt}

    \textbf{security} \dotfill\ {\small hardware-based/unbounded network adversary}\\
    \textbf{set-up} \dotfill\ {\small enrolled physical hardware}\\
    \textbf{composability} \dotfill\ open\\
    \textbf{public verifiability} \dotfill\ N/A\\
    \textbf{scalability} \dotfill\ {\small linear and device-dependent}\\
    \textbf{candidate construction} \dotfill\ {\small HLPUF entity authentication}
  };

  \node[
    draw,
    rounded corners=3pt,
    fill=white,
    font=\bfseries,
    inner xsep=6pt,
    inner ysep=2pt
  ] at ([xshift=-0.2\linewidth]box.north) {EA};

\end{tikzpicture}


\section{Applications and Deployments}\label{sec:apps-and-depls}
Now that we have seen various methods of achieving various modes of authentication, let us see some of them in action.

\subsection{Authentication for QKD}\label{sec:auth-for-qkd}
QKD is easily the most actively deployed, and probably most studied, quantum cryptography protocol. Let us now see what kind of security profile can be achieved when QKD is used in tandem with various modes of classical message authentication, and review the appearance of these combinations in the literature.

Two previous review-type papers have already investigated QKD from a similar perspective. In accordance with our thesis that it is necessary to investigate authentication mechanisms for use on the classical channel resource of QKD, All{\'e}aume et al \cite{alleaume_et_al_cryptographic_purposes} discuss the issues with requiring the distribution of pre-shared keys in large networks, and argue that QKD and post-quantum cryptography can be regarded as complementary technologies to QKD in the establishment of quantum-safe networks. In particular, we echo this group's characterisation of a system combining QKD with PQC authentication -- that is, as a quantum-safe key exchange mechanism granting post-quantum computational security during run-time, long-term unconditional security, and scalable authentication.

In discussing this characterisation of the combination of QKD and PQC, the survey of All{\'e}aume et al makes reference to a discussion of Paterson, Piper and Schack \cite{paterson_skepticism}\footnote{We also note that a response continuing the discussion induced by this work \cite{the_case_for_qkd} has some overlap with our discussion.}, which first appeared on the ePrint archive in 2004 -- more than 20 years ago. This is the second of the two discussion papers close to the spirit of the present work. Interestingly, the authors of that paper conclude that, since they do not believe that the long-term security of practical, computationally secure symmetric cryptography is under threat, the practical response to the advent of cryptographically relevant quantum computers should not be to combine the two complementary `quantum-safe' technologies of QKD and PQC, but to re-establish the use of symmetric key techniques. As an example, they cite the use, in mobile networks, of SIMs equipped with 128-bit private keys, which in mid-2003\footnote{At the time, a contemporaneous example.} saw extremely widespread deployment. 

With the benefit of hindsight, we now know that this point of view did not enter the mainstream (although certainly it has its advocates). Nevertheless, it is a helpful illustration of the need for this kind of survey; that is, a survey taking into account a more complete spectrum of approaches and outlooks to authentication in quantum cryptography. More prosaically, since the survey of All{\`e}aume et al appeared in 2014, there are simply many more examples of papers proposing to use post-quantum cryptography in tandem with quantum cryptography.

\subsubsection{Schemes using symmetric authentication}

\paragraph{Protocols.} In general, most papers discussing instantiation of QKD are at least implicitly suggesting that symmetric key authentication be used with their protocol. The most famous examples of QKD protocols either explicitly suggest this method; mention the need for an authenticated channel, without specifying the authentication mechanism; or simply do not discuss authentication (put another way, they only consider eavesdropping on the classical channel). For example, BB84 \cite{bb84_original}, B92 \cite{b92}, the high-rate DI protocol of \cite{high-rate-mdi}, and the local Bell test variant of a device-independent protocol \cite{DI_local_bell_test} fall into this first camp\footnote{Interestingly, the latter does not do so directly, but appeals to composability.}; the no-signaling protocol of Barrett, Hardy and Kent falls into the second camp \cite{no-signaling_protocol}; and E91 \cite{e91}, the six-state protocol \cite{six-state}, the original continuous-variable protocol \cite{og-cv}, BBM92 \cite{bbm92}, MSZ96 \cite{msz96}, and the differential phase shift protocol of Inoue, Waks and Yamamoto \cite{inoue_phase_shift} all fall into the latter. An outlier in this sense is a study on post-processing more broadly with respect to QKD\cite{eval-of-different-sk}, which offers a thoughtful analysis of the suitability of various different symmetric-key protocols.

\paragraph{Security proofs.} There are sufficiently many proofs of security of quantum cryptography, across different protocols and techniques\footnote{As the field has evolved, different measures of the information an eavesdropper can obtain on the quantum channel have been recognised as appropriate, and proofs have become more sophisticated in order to formally capture desirable properties of quantum cryptography protocols. Again, the details here are out of scope for our purposes.}, as to require a survey on the matter \cite{qkd_security_review}. This type of paper fits in one of three categories: one does not mention authentication at all, or as we have already discussed, can be said to prove security against eavesdroppers or passive adversaries. Examples of this type of paper include \cite{proof-cat-1-ex-1, proof-cat-1-ex-2, proof-cat-1-ex-3, proof-cat-1-ex-4, proof-cat-1-ex-5, proof-cat-1-ex-6, proof-cat-1-ex-7, proof-cat-1-ex-8, proof-cat-1-ex-9, proof-cat-1-ex-10}. Some other proof-type papers mention the need for a classical authenticated channel, but do not specify the mechanism by which this should be realised: for example, \cite{proof-cat-2-ex-1, proof-cat-2-ex-2, proof-cat-2-ex-3, proof-cat-2-ex-4, proof-cat-2-ex-5, proof-cat-2-ex-6, proof-cat-2-ex-7}. Finally, there are papers that explicitly mention that the authenticated channel should be realised by means of short, pre-distributed keys, including \cite{proof-cat-3-ex-1, proof-cat-3-ex-2, proof-cat-3-ex-3, proof-cat-3-ex-4, proof-cat-3-ex-5,proof-cat-3-ex-6, proof-cat-3-ex-7, proof-cat-3-ex-8, proof-cat-3-ex-9}. We mention as an outlier that one paper we surveyed justified their assumption that an insecure channel could be trusted thusly: ``If Alice and
Bob could not afford to receive any untrusted classical message, the whole enterprise of
cryptography would be hopeless'', which appears to be a misunderstanding of the assertion that any authenticity guarantee requires some degree of trust. 

\paragraph{Implementation and deployment.} Finally, let us review the means of authentication suggested in the practical deployment of QKD. The first demonstration of QKD \cite{first-practical-qkd} distilled a key of 175 bits over a distance of 30cm, so in this case the researchers were confident of securing their classical transmission by whispering. Later, improvements to key rates were made by the same group of researchers \cite{second-practical-qkd}, but this set-up was also over a short distance within a lab, and so the integration of an authentication mechanism was not discussed. 

In general (and with a few exceptions), we found that the papers reporting on the practicalities of QKD do not contain satisfactory discussion of how the protocol would be authenticated. This is understandable in a lab-based context, when some improvement to range, or key-rate is being reported on, and the interest is more in some improvement to technology, rather than describing a full-stack communication solution. Examples of this kind of paper include \cite{long-distance-mdi-demo, exp-no-mention}, which do not mention the need for classical authentication; \cite{extra-cable-1, extra-cable-2}, which only mention needing additional fibre dedicated to classical communication; and \cite{exp-mentions-auth, T12-protocol}, which mention the need for a classical authenticated channel, without suggesting a means by which this should be achieved. 

In our view, the omission of authentication is more troubling in papers reporting on deployment \textit{in situ} -- say, as a prototype for some metropolitan area QKD network, or as a proposed commercial system -- because these real-life use cases would require some means of authentication in practice. Here, the literature is quite varied: some make no mention of an authenticated channel \cite{depl-no-mention-1, toshiba-post, toshiba-post-2}; some mention their classical channel is authenticated with pre-distributed keys \cite{bristol-trial, tokyo-trial}, or at least acknowledge the need for such a channel \cite{madrid-trial}. We mention a particularly illustrative outlier\cite{depl-error-2}. Roughly speaking, this paper wishes to achieve information-theoretic classical authentication by using keys derived from QKD. As already pointed out\cite{use-case-list}, the method of authentication suggested is not actually information-theoretically secure; even worse, there is no mention at all of how one would construct a classically authenticated channel to construct this QKD link. In other words, the authors have the situation almost exactly backwards: rather than discuss the information-theoretically authentic classical channel that a QKD link must be constructed from in order to yield information-theoretically secure keys, the authors treat QKD as a black-box, from which keys can be derived to achieve information-theoretically secure authentication.

Finally, we note a recent exception to this trend in the literature: a QKD network across a metropolitan area has been reported with careful attention to the classical authentication, which is explicitly implemented and benchmarked\cite{sym-keys-in-the-city}. 

Let us turn our attention to the authentication of QKD with public-key cryptography. Note that we continue to rather pointedly use the term `public-key' (as opposed to`post-quantum') cryptography, for reasons that will appear in the discussion.

\paragraph{Evaluation}
Because unconditional authentication of a message requires keys of length roughly logarithmic in that message, we can in theory hope to obtain more key material than we started with if we use universal hash functions to authenticate the classical transmission in a QKD protocol. Because of this property QKD is sometimes called quantum key \textit{expansion}; in other words, it is theoretically possible from a small, pre-shared secret key to obtain an unlimited amount of unconditionally secure (i.e. perfectly secret) key material. Notice that this is not possible classically, because of Shannon's theorem: a fresh session key requires a pre-shared key of equal length to ensure its perfect secrecy.

In practice the quantum bit error rate might be such that the expansion effect is not sufficient for the application, in which case it can be increased by relaxing the amount of key required even further. This is possible using message-authentication codes and provides real-time security assuming the existence of one-way functions, and unconditional long-term security. On the other hand, one has precisely the same scaling profile as for universal hash functions (i.e. one-time distribution of pre-shared keys), and if one is willing to tolerate security from one-way functions, more scalable options are available from public-key cryptography.

\subsubsection{Schemes using public-key authentication}
Recall our composable view of authentication: a QKD protocol is composed of the distribution of some quantum information and communication on an authenticated channel. In order to authenticate QKD with PKC, then, one need only make changes to the way in which the authenticated channel is constructed -- the `quantum part' is the same.

\paragraph{Modes of hybridisation.} We briefly remark that there is something of a split in the literature on protocols which use both QKD and PKC. On the one hand, there are protocols which are interested only in PKC as the authentication mechanism for a standard QKD protocol; recently, however, several papers offering a `defence-in-depth' approach have appeared, whereby a key is derived from secrets derived both from QKD and from PKC\cite{muckle,muckle-plus,muckle-sharp,muckle-plus-plus,vmuckle,qsoft-group,non-block-box-qkd, more-non-black-box}. Henceforth we focus on the former mode. 

\paragraph{Protocols}
The canonical example here is the protocol of Mosca, Stebila and Ustao{\u{g}}lu\cite{mosca-protocol}, where signatures are used for authentication and security is proved in a classical authenticated-key exchange framework. Various improvements and additional security features appear subsequently: for example, reduced communication cost\cite{pqc-auth-1} and identity protection\cite{pqc-auth-2}, benchmarking of different schemes\cite{pqc-auth-benchmark}, and composable security \cite{pqc-auth-composable-1,pqc-auth-composable-2}. 

\paragraph{Implementation and Deployment}
A number of papers describing deployment of public-key authentication of QKD have appeared in the literature recently. This line of work was pioneered by Wang et al\cite{pqc-exp-1}; later, hybrid symmetric-key/public-key authentication was demonstrated\cite{pqc-exp-2}. The approach has also been demonstrated in larger networks\cite{pqc-exp-3,pqc-exp-4,pqc-exp-5}, which is of particular relevance given the scalability advantages offered by this approach.

\paragraph{Evaluation}
All the public-key protocols we have described offer everlasting security; because of no-cloning, their compromise after run-time does not affect the long-term unconditional security of any established keys. The question then is which mechanism to use; that is, KEM-then-MAC, Lamport-style signatures, or post-quantum signatures based on trapdoor functions\footnote{One could in theory advocate for authenticating QKD with quantum-vulnerable signatures in a particular threat model whereby one believes that quantum computers will not cause a threat to classical cryptography, but one is worried about potential classical cryptanalytic advances in the long-term.}? 

KEM-then-MAC authentication offers a significant save on bandwidth in the post-quantum context, but this is unlikely to be an efficiency bottleneck in the short-to-medium term, as quantum devices currently dominate QKD runtime\cite{pqc-exp-1}. In the long term the approach may be desirable in resource-constrained environments. One also has to rely on the existence of quantum-resistant trapdoor functions during runtime. The same is true of post-quantum signatures based on trapdoor functions.

Lamport-style signatures are of particular interest here, because their security assumes only the existence of one-way functions. This effectively collapses the assumed classical hierarchy: it is generally believed that it is not possible to achieve key agreement without pre-shared keys in the classical setting, except via constructions based on trapdoor functions. By authenticating QKD with signatures constructed from one-way functions we can achieve key agreement from one-way functions. Although the Lamport-style constructions are extremely inefficient compared to their trapdoor-based counterparts, this is unlikely to be an efficiency bottleneck in the short term.

\subsubsection{Schemes using hardware assumptions}
Hardware-based authentication protocols are mostly designed for entity authentication rather than message authentication, due to their challenge-response nature, as discussed in detail in Section~\ref{sec:identification}. For this reason, they are not immediately ready to replace the classical message-authentication layer required in QKD. A recent line of work has nevertheless explored how hardware tokens can be used to generate keys for standard classical message authentication, effectively by using the hardware responses as authentication-key material. This approach was initiated by Nikolopoulos and Fischlin~\cite{qkd-with-pufs} and has since been adopted in several QKD-oriented works~\cite{qkd-wtih-pufs-2,qkd-with-pufs-3}.

The security interpretation of these schemes requires some care, as most of these schemes are built based on classical PUFs, and thus the resulting protocol inherits the usual limitations of classical PUF-based security, and does not necessarily provide the full quantum security required for QKD. Thus, while these protocols provide an interesting hardware-based route to bootstrapping authentication keys for QKD and might be interesting practical solutions, their quantum-security guarantees should be understood as conditional on the concrete PUF assumptions being satisfied, and often still valid in some heuristic regime.

Very recently, this direction has been strengthened by schemes based on quantum-secure hybrid hardware assumptions. In particular, the hybrid entangled PUF approach of Ref.~\cite{qkd-with-pufs-4} integrates hardware-based authentication directly with entanglement-based QKD. The protocol uses a hybrid PUF authentication subroutine to generate an initial secret key, which is then used to authenticate the subsequent classical communication in the QKD protocol through a standard information-theoretically secure message-authentication scheme. This gives, to the best of our knowledge, the first fully provable demonstration of hardware-based authentication integrated with QKD against unbounded quantum-channel adversaries, under explicit hybrid hardware assumptions. Related self-authenticated key-exchange schemes based on hybrid hardware assumptions have also appeared in the patent literature~\cite{qpufpatent2025}. Experimental deployments of these schemes report favourable key rates

From a security perspective, the main advantage compared to pre-shared keys is that compromise of the hardware token need not immediately void all security guarantees, provided the adversary remains within the stated hardware assumptions. By contrast, pre-shared classical keys cannot tolerate any compromise of the key material itself. One should nevertheless be careful about claims of ``unconditional security'' in this setting. Such claims apply to the message-authentication stage once the required set-up phase has been completed and the initial key has been established. Moreover, the choice of the subsequent classical authentication scheme is critical: the established key yields unconditional authentication only when it is used within an unconditionally secure classical authentication scheme. In this sense, the construction is analogous to a KEM-then-MAC approach, where the output key of the key-establishment mechanism is then used as the key in a classical authentication protocol.

We also comment on the scaling of these approaches. Papers in this area often emphasise favourable scaling in large networks compared to pre-shared keys, and there is a meaningful sense in which hardware tokens can reduce the operational burden of authentication. The situation is nevertheless subtle. On the one hand, in the most direct deployment model, each pair of honest parties still needs access to a distinct hardware token or hardware-derived secret, and a new participant may need to register separately with every existing member of the network. On the other hand, the ability to tolerate some forms of hardware-token compromise can make the registration and distribution process more flexible than in the case of ordinary pre-shared keys, potentially reducing the level of trust required in couriers or provisioning infrastructure. Thus, hardware-based schemes do not automatically remove pairwise set-up costs, but they can change the operational and security profile of that set-up in a way that may be advantageous for quantum-network deployments.

\subsection{Advanced functionalities beyond QKD}\label{sec:qma-apps}

Compared with QKD, the literature on authentication in more advanced
quantum-cryptographic protocols is less mature and much less tied to
deployment. The relevant question is also different. In QKD, the main
issue is how to instantiate the authenticated classical channel. Beyond
QKD, authentication can appear in several places: as integrity for a
classical transcript, as protection of quantum data, as certification of
a party, device, or location, or as part of the proof technique used for
verification. This is exactly the distinction captured by the three
functionalities used in this review.

\begin{itemize}
    \item \textbf{Classical message authentication} is the common
    baseline. Protocols such as quantum anonymous
    transmission~\cite{quantum-anon-transm}, Byzantine
    agreement~\cite{byzantine-agreement}, zero-knowledge
    proofs~\cite{everlasting-zk-qma}, and anonymous conference key
    agreement~\cite{hahn2020anonymous,grasselli2022anonymous} use
    classical communication whose integrity is part of the security
    model\footnote{For a helpful map of the relationship between
    functionalities and resources, see the
    \href{https://wiki.veriqloud.fr/index.php?title=Graphs}{Quantum
    Protocol Zoo}.}. The anonymous-conferencing protocols make this
    explicit by naming authenticated broadcast and pairwise
    authenticated channels among their required resources. Relativistic
    bit commitment~\cite{lunghi2015practical} is a concrete
    implementation-level case: the security of the unveiling phase rests
    on an information-theoretic message-authentication code applied to
    communication between the agents. Anonymous transmission is a useful
    edge case. The broadcast transcript must have integrity, but the
    sender's public identity must remain hidden. The relevant binding is
    therefore to membership in the authorised network and to the
    integrity of the transcript, not to a revealed sender identity.

    \item \textbf{Quantum message authentication} is relevant when
    quantum data itself must be protected against tampering. Early
    quantum digital signature schemes needed quantum authentication to
    distribute quantum public keys, although later variants removed this
    requirement, as discussed in Section~\ref{sec:qdss}. Quantum
    fingerprinting~\cite{q-fingerprinting} is another example where an
    authenticated quantum channel may be required. The clearest modern
    use is secure quantum computation. In secure two-party and
    multi-party quantum computation~\cite{dupuis2012actively,
    dulek2020secure,kapourniotis2025asymmetric}, parties compute on
    authenticated quantum data so that deviations are detected rather
    than silently propagated. Quantum one-time
    programs~\cite{broadbent2013onetime} provide another explicit
    example: the program is encoded using a quantum authentication
    scheme, and evaluation proceeds over authenticated data.

    \item \textbf{Entity authentication} is the relevant set-up primitive
    when a party, device, or location must be certified before the rest
    of the protocol has operational meaning. The hardware-based
    protocols of Section~\ref{sec:identification} are natural candidates
    for this role in quantum-network deployments. Position-based quantum
    cryptography is a related variant: the credential is not an
    intrinsic identity, but a claimed spatial location. In models where
    secure position verification is achievable~\cite{buhrman2014position},
    it can support position-based authentication and key agreement \footnote{This use of position as a credential must be qualified: with arbitrary pre-shared entanglement between adversaries, position verification is impossible. The authentication role is meaningful only in models that restrict such entanglement.}.

    \item \textbf{Verifiability and Integrity:} Verifiable blind quantum computation (VBQC) provides a useful comparison
point. In UBQC and VBQC~\cite{broadbent2009universal,
fitzsimons2017unconditionally,blind-qc-overview}, a client delegates a
computation to an untrusted server while hiding the input, output, and
computation, and in the verifiable case detecting an incorrect result
except with bounded probability. The classical transcript, including
measurement instructions and feed-forward outcomes, needs integrity in
any network implementation. This can be supplied by an authenticated
classical channel, or treated as part of the adversarial behaviour of
the server in the security model. Entity authentication is a separate
deployment issue: it is needed when the client must bind the execution
to a particular provider, device, or network node, but it is not the
primitive that gives blindness or verifiability. The transmitted qubits
are not usually protected by an independent quantum-message-authentication
layer. Their disturbance is handled by the blindness and verification
mechanisms of the delegated-computation protocol, for example through
hidden traps or authenticated-code structure in protocols that use such
codes. VBQC therefore illustrates how message integrity, identity
binding, and verification of delegated computation interact, rather than
making all three authentication functionalities explicit as separate
resources.

The role of authentication is not limited to that of an external
resource. In some delegated-computation protocols, authentication is
part of the proof technique itself. In authentication-code and trap-based
verification protocols, encryption, randomisation, or blinding reduces
attacks to deviations that can be analysed through Pauli errors or
error-detection tests. Some verification protocols are built directly
from quantum authentication schemes~\cite{aharonov2010interactive,
broadbent2018how}. More generally, the relation between verification
and error detection can be made explicit by viewing traps as
probabilistic error-detection tests~\cite{kapourniotis2024unifying,
gheorghiu2019verification}.

A key ingredient that lets authentication support verification is
controlled computation on protected data. This is not malleability in
the adversarial sense discussed in Section~\ref{sec:props-of-auth}. The
allowed transformations are part of the protocol specification: they
preserve the authentication structure, or come with updates to secret
keys, traps, syndromes, or Pauli frames. Operations outside this
interface must still be detected, or reduce to an error model caught by
the final tests.

Authentication codes that support gates through the encoding, for
instance Clifford-transversal codes supplemented by gadgets for
non-Clifford operations, allow a delegated computation to run while
keeping the authentication keys hidden. This idea appears in secure
two-party quantum computation~\cite{dupuis2010secure,dupuis2012actively},
in secure multi-party quantum computation~\cite{dulek2020secure}, in
quantum one-time programs~\cite{broadbent2013onetime}, and in
verifiable quantum homomorphic encryption~\cite{alagic2017qfhe}. It
mirrors the classical problem of computing on authenticated
data~\cite{ahn2012computing,gennaro2013fully}. A general
characterisation of which quantum-authentication schemes can be
converted into composable verification protocols remains open, although
sufficient conditions are known for important trap-based and
error-detection-based families.

A future case study should therefore ask not only whether authentication
is assumed, but which authentication resource is assumed, whether it is
message-level, quantum-data-level, or entity-level, and whether the
assumption is implementable in the proposed network model.
\end{itemize}

\section{Outlook}\label{sec:outlook}
We have seen that the term `authentication' encompasses a wide variety of functionalities, and the means of realising these functionalities have various advantages and drawbacks. Our biggest take-home message is that one cannot hope for a perfect solution: as we have seen, the objectives of assumption-free security and practicality generally conflict, and so one must carefully balance the strength of security that can be tolerated, compared to the ability to properly calibrate the system. The conflicting incentives within classical message authentication are particularly complex; to highlight the decision-making progress refer to Figure~\ref{fig:auth-flow}. We highlight two subtle points of interest within the diagram: QKD, by its expansion property, allows one to repeatedly establish keys, and therefore allows one to use UHFs even from one-time key distribution; and the fact that cryptomania is at least as strong an assumption as minicrypt is reflected in the order that questions are asked in.

The picture is less clear for quantum message authentication and entity authentication. Unlike in the case of message authentication, these functionalities can be considered the applications themselves; that is, they are of intrinsic interest from a quantum networks perspective without being considered as input to a more advanced protocol. Indeed, since quantum authentication needs pre-shared keys, or authentic copies of public keys, one can view classical message authentication as a necessary input for quantum message authentication. On the other hand, there do exist protocols requiring quantum authentication and entity authentication as input, though they are much sparser than those requiring classical message authentication as an input. One might expect this to change as the field matures.  

Across all three functionalities, the outlook looks optimistic in terms of composability. For most of the classical authentication primitives we have mentioned the composability is understood\cite{composable-review}; protocols for the other functionalities are more explicitly quantum, where the literature tends to define security in a simulation-based (and therefore at least plausibly composable) fashion. Nevertheless, subtle questions remain pertaining to composability in multi-party settings, and so on. 

We have paid close attention to scalability throughout the review, defining the cost of joining a network in the worst case as our proxy metric. A more careful analysis of scalability in different, non-worst-case network architectures would also be interesting: for example, the asymmetric memory requirements of some hardware-based protocols suggest that they are particularly suitable for star-shaped networks.

All of our analysis has been qualitative. Further work benchmarking and comparing the explicit performances of these protocols would be a welcome complement to this work, particularly if deployed in real test-beds.

We opened this review with the observation, made by several national security agencies~\cite{nsa_position}, that QKD does not by itself authenticate its transmission source. The broader lesson of the survey is that this is not special to QKD. Every quantum cryptographic functionality we have considered, from key distribution to delegated computation, rests at its base on an authentication assumption, whether a pre-shared key, an authentic copy of a public key, or an enrolled piece of hardware, as set out in Section~\ref{sec:framework}. Authentication is in this sense the root of trust for distributed quantum protocols, and the security guarantee of any such protocol is no stronger than the authentication assumption beneath it. This dependence is easy to overlook, since the authentication step is so often stated only implicitly, as both our review of QKD deployments (Section~\ref{sec:auth-for-qkd}) and our discussion of the advanced functionalities have shown. As quantum infrastructure grows from point-to-point QKD links towards networked computation, sensing, and multi-party tasks, the central task is therefore not only to design new mechanisms, but to state precisely which authentication assumption each protocol requires and to establish whether that assumption can be met in deployment. Clarifying this root of trust, and the gap between the assumptions made in security proofs and those that can be realised in practice, is in our view a precondition for trustworthy distributed quantum infrastructure.\\

\noindent \textbf{Acknowledgment}\\
The authors thank Myrto Arapinis for valuable discussions and comments during different stages of this review. CB, MD and EK acknowledge the support of the Integrated Quantum Networks Hub, grant reference EP/Z533208/1. MD, SG and EK also acknowledge the support of Quantum Advantage Pathfinder (QAP), with grant reference EP/X026167/1. MD and EK acknowledge Quantum Advantage Turbo Charger (QATCH) with grant reference UKRI4257.

\bibliographystyle{naturemag}
\bibliography{references}

\end{multicols}

\end{document}